\documentclass[journal]{IEEEtran}

\ifCLASSINFOpdf
   \usepackage[pdftex]{graphicx}
\else
   \usepackage[dvips]{graphicx}
   
  \graphicspath{{../eps/}}
 
\fi

\usepackage{dblfloatfix}
\usepackage{xcolor,soul,framed} 
\usepackage{xcolor}
\colorlet{shadecolor}{yellow}
\usepackage{changes}
\usepackage[linesnumbered,ruled,vlined]{algorithm2e}
\usepackage{caption}
\usepackage{subfig}
\usepackage[cmex10]{amsmath}
\usepackage{array}
\usepackage{mdwmath}

\usepackage{amssymb}
\usepackage{mdwtab}
\usepackage{amsfonts}
\usepackage{eqparbox}
\usepackage{url}
\usepackage{balance}

\usepackage{cite}
\graphicspath{{../pdf/}{../jpeg/}}
\DeclareGraphicsExtensions{.pdf,.jpeg,.png}
\usepackage{acro} 
\DeclareAcronym{RIS}{
  short = RIS ,
  long  = reconfigurable intelligent surface ,
  class = abbrev
}
\DeclareAcronym{6G}{
  short = 6G,
  long  = sixth-generation ,
  class = abbrev
}
\DeclareAcronym{EM}{
  short = EM ,
  long  = electromagnetic,
  class = abbrev
}
\DeclareAcronym{BS}{
  short = BS,
  long  = base station ,
  class = abbrev
}
\DeclareAcronym{UE}{
  short = UE ,
  long  = user equipment ,
  class = abbrev
}
\DeclareAcronym{MISO}{
  short = MISO,
  long  = multiple-input single-output ,
  class = abbrev
}
\DeclareAcronym{MMSE}{
  short = MMSE ,
  long  = minimum mean squared error ,
  class = abbrev
}
\DeclareAcronym{DFT}{
  short = DFT,
  long  = discrete Fourier transform ,
  class = abbrev
}
\DeclareAcronym{THz}{
  short = THz,
  long  = Terahertz ,
  class = abbrev
}
\DeclareAcronym{IoT}{
  short = IoT,
  long  = internet of things ,
  class = abbrev
}
\DeclareAcronym{MSE}{
  short = MSE,
  long  = mean square error ,
  class = abbrev
}
\DeclareAcronym{CSI}{
  short = CSI ,
  long  = channel state information ,
  class = abbrev
}
\DeclareAcronym{MIMO}{
  short = MIMO,
  long  = multiple-input multiple-output ,
  class = abbrev
}
\DeclareAcronym{UPA}{
  short = UPA,
  long  = uniform planner array ,
  class = abbrev
}
\DeclareAcronym{RF}{
  short = RF,
  long  = radio-frequency ,
  class = abbrev
}
\DeclareAcronym{mmWave}{
  short = mmWave,
  long  = millimeter-wave ,
  class = abbrev
}
\DeclareAcronym{AoA}{
  short = AoA ,
  long  = angle of arrival ,
  class = abbrev
}
\DeclareAcronym{AoD}{
  short = AoD,
  long  = angle of departure ,
  class = abbrev
}
\DeclareAcronym{EKF}{
  short = EKF,
  long  = extended Kalman filter ,
  class = abbrev
}
\DeclareAcronym{LMS}{
  short = LMS,
  long  = least mean square ,
  class = abbrev
}
\DeclareAcronym{BiLMS}{
  short = BiLMS,
  long  = bi-directional LMS ,
  class = abbrev
}
\DeclareAcronym{SNR}{
  short = SNR,
  long  = signal-to-noise ratio ,
  class = abbrev
}
\DeclareAcronym{LoS}{
  short = LoS,
  long  = line-of-sight ,
  class = abbrev
}
\DeclareAcronym{TDD}{
  short = TDD,
  long  = time-division duplexing ,
  class = abbrev
}
\DeclareAcronym{NMSE}{
  short = NMSE,
  long  = normalized mean square error ,
  class = abbrev
}
\DeclareAcronym{2D}{
  short = 2D,
  long  = two-dimensional ,
  class = abbrev
}
\DeclareAcronym{3D}{
  short = 3D,
  long  = three-dimensional ,
  class = abbrev
}

\begin{document}
\setlength{\marginparwidth}{2cm}
\bstctlcite{IEEEexample:BSTcontrol}
    \title{A General Framework for RIS-Aided mmWave Communication Networks: Channel Estimation and Mobile User Tracking}

\author{
    \IEEEauthorblockN{Salah Eddine Zegrar,}
    \IEEEauthorblockN{Liza Afeef,} 
	\IEEEauthorblockN{and}
	\IEEEauthorblockN{H\"{u}seyin Arslan,}
	\IEEEmembership{Fellow, IEEE}
	
\thanks{The authors are with the Department of Electrical and Electronics Engineering, Istanbul Medipol University, Istanbul, 34810, Turkey (e-mail: salah.zegrar@std.medipol.edu.tr; liza.shehab@std.medipol.edu.tr;  huseyinarslan@medipol.edu.tr).}

\thanks{H. Arslan is also with Department of Electrical Engineering,
	University of South Florida, Tampa, FL, 33620, USA.}

\thanks{This work has been submitted to the IEEE for possible publication. Copyright may be transferred without notice, after which this version may no longer be accessible.}
}

\maketitle

\begin{abstract}
Reconfigurable intelligent surface (RIS) has been widely discussed as new technology to improve wireless communication performance. Based on the unique design of RIS, its elements can reflect, refract, absorb, or focus the incoming waves toward any desired direction. These functionalities turned out to be a major solution to overcome millimeter-wave (mmWave)’s high propagation conditions including path attenuation and blockage. However, channel estimation in RIS-aided communication is still a major concern due to the passive nature of RIS elements, and estimation overhead that arises with multiple-input multiple-output (MIMO) system. As a consequence, user tracking has not been analyzed yet. This paper is the first work that addresses channel estimation, beamforming, and user tracking under practical mmWave RIS-MIMO systems. By providing the mathematical relation of RIS design with a MIMO system, a three-stage framework is presented. Starting with estimating the channel between a base station (BS) and RIS using hierarchical beam searching, followed by estimating the channel between RIS and user using an iterative resolution algorithm. Lastly, a popular tracking algorithm is employed to track channel parameters between the RIS and the user. System analysis demonstrates the robustness and the effectiveness of the proposed framework in real-time scenarios. 
\end{abstract}

\IEEEpeerreviewmaketitle

\section{Introduction}

\IEEEPARstart{M}{illimeter-wave} (mmWave) communication has become one of the key technologies of fifth-generation (5G) communication systems. Although mmWave achieves high data rate due to its wider signal bandwidth, it suffers from severe path loss\cite{mmwaves-pathloss}. Many solutions can be implemented to overcome these losses including high-dimensional \ac{MIMO} operations, and \ac{RIS} technology to mitigate the limitation conditions of high frequencies.
Recently, \ac{RIS} has attracted much attention as a highly promising technology that can meet the requirements of the \ac{6G} and beyond wireless networks. \ac{RIS}’s capability arises from its ability to support \ac{MIMO} systems in controlling and hardening the wireless channel, where a highly time-varying channel can behave as a deterministic one. The functionalities of \ac{RIS} including reflecting, diffracting, or scattering the transmitted signal enhance the quality of the signal at the receiver side. 
These abilities come from its unique design where the adjustable passive elements can individually steer the incident \ac{EM} wave toward any specific direction by changing their phases and gains only. Adjusting these elements allows us to align all multipath of the reflected signal so that they are added constructively at the receiver \cite{1}.
This principle of the \ac{RIS} elements fulfills the concepts of beamforming and steering concepts \cite{13,emilMyths}. Therefore, with proper \ac{RIS} size and reflection coefficients, the reflected signal is a beam, where the width of this beam is inversely proportional to the size of the \ac{RIS}. Since these elements passively reflect the signal, they are easy to implement, have a low-cost deployment, and most importantly do not cause noise amplification \cite{2}. These features makes the \ac{RIS} a strong candidate in upcoming wireless systems over conventional technologies.

On the other side, \ac{RIS} imposes a lot of challenges such as channel estimation. Since the \ac{RIS} is built of a large number of passive elements, \ac{RIS}-aided communication networks have faced difficulties in estimating the channel reliably. To overcome these difficulties, many channel estimation techniques have been proposed in the literature under different approaches.

In single-user systems, \cite{3} proposes a novel \ac{RIS} architecture in which some of its elements are active by connecting them to a baseband processor. These active elements turned back to the reflecting mode after estimating the channel. However, this sensor deployment, that activates the elements, increases the implementation cost. Prior works \cite{4,6,9,Alouini2,8,5,7,11} focused on introducing channel estimation techniques with fully passive \ac{RIS} elements. A two-stage algorithm, in \cite{4}, for channel estimation is proposed. In the first stage, all RIS elements are turned off whereas the direct channel between the \ac{BS} and \ac{UE} is estimated. Then, in the second stage, the elements are turned on one by one while the channel between each element and \ac{UE} is being estimated. In spite of that, this strategy degrades the channel estimation accuracy since only a small portion of \ac{RIS} elements are switched on at each time. This also requires using a separate amplitude control and phase shifter for each element which increases the system cost, especially for a massive number of \ac{RIS} elements. Therefore, the approach in \cite{6} divides the \ac{RIS} elements $N$ into $M$ sub-surfaces while keeping the elements on with maximum reflected amplitude during the channel estimation and data transmission. Each sub-surface consists of $N/M$ adjacent elements with a common reflection coefficient to reduce the implementation complexity. 
Following the same approach, \cite{9} proposes a \ac{DFT}-based channel estimation method where all the \ac{RIS} elements are on at all time slots and the \ac{DFT} matrix is used to determine the reflection coefficients of these elements. In addition, \ac{MMSE} algorithm is proposed in \cite{Alouini2} to work with \ac{DFT} matrix to estimate the channel of both direct path and the \ac{RIS}-assisted path between the \ac{BS} and \ac{UE} in \ac{MISO} system. 
For more realistic setting, \cite{8} considers discrete-phase \ac{RIS} reflecting elements that are grouped into a relatively small number of sub-surfaces, and proposes a \ac{DFT}-Hadamard-based reflection pattern strategy to minimize the channel estimation error in imperfect \ac{CSI}. However, increasing the number of sub-surfaces causes training overhead and degradation in the spectrum efficiency. Meanwhile, free-space path loss models for the \ac{RIS}-assisted wireless communication are proposed in \cite{13,12} using the \ac{EM} and physical properties of a reconfigurable surface. In \cite{13}, a \ac{2D} path loss model is derived and extended to \ac{3D} one in \cite{12} including far-field, near-field beamforming, and near-field broadcasting formulas are experimentally validated in the indoor environment.

In a multiuser system, \cite{5} investigates the \ac{MISO} system under imperfect CSI, where it considers correlated Rayleigh fading channel. The proposed protocol utilizes \ac{DFT}-\ac{MMSE} estimation and divides channel estimation into subphases, at each one, all users transmit orthogonal pilot symbols to estimate the \ac{RIS}-assisted links.
The work in \cite{7} proposes a three-phase channel estimation framework for the \ac{RIS}-assisted uplink \ac{MISO} system to reduce the training duration. In phase one, \ac{RIS} elements are turned off and direct channels between \ac{UE}s and \ac{BS} are estimated. Then, in the next phase, among all \ac{UE}s, only one \ac{UE} transmitted pilots and the cascaded channel is estimated. In the last phase, the channel between all \ac{UE}s and the \ac{RIS} is considered correlated, thus only the scaling factors need to be estimated. However, the training overhead, the number of supportable users, and the performance of channel estimation are the limits of this technique.

In mmWave systems, channel estimation becomes more critical with few works touching this problem \cite{3,estimation-mm,estimation-mm2,11}. The work in \cite{11} proposes a compressed-sensing-based channel estimation algorithm where the sparsity of the channel is exploited to implement channel estimation at a reduced pilot overhead in massive \ac{MIMO} system. The sparsity of the channel is capitalized using a distributed orthogonal matching pursuit algorithm. It is assumed that there is prior knowledge about the channel between the \ac{BS} and the \ac{RIS}, and the pilots are designed accordingly. However, considering the channel \ac{BS}-\ac{RIS} to be known and time-invariant are not practical since mmWave channel is sensitive to small changes. 
Similarly, a compressed sensing algorithm is utilized in \cite{compressed-sensing1} to estimate the cascaded channel parameters in the \ac{RIS}-assisted THz \ac{MIMO} system in the indoor application scenarios, since it is assumed that the estimation problem is equivalent to sparse recovery problem. Again, using the same method, \cite{compressed-sensing2} tries to find a sparse representation of the cascaded \ac{BS}-\ac{RIS}-\ac{UE} channel in mmWave downlink system with the help of transposed Khari-Rao product and Kronecker product. The most recent work for channel estimation in mmWave systems was in \cite{High-resolution}, where a two-stage cascaded channel estimation protocol is proposed by exploiting the sparsity of mmWave \ac{MIMO} channel of single \ac{BS}, \ac{RIS}, and \ac{UE}. In the first stage, the beam searching approach is introduced to have high angular domain information, then in a second stage, an adaptive grid matching pursuit algorithm is proposed to estimate the high-resolution cascaded channel. However, estimating a cascaded channel has many limitations as it will be explained and proved later in this paper. 

Although the aforementioned channel estimation techniques are theoretically effective with low \ac{MSE} level, they depend on either cascaded channel concept or non-practical assumptions for estimating the channel \ac{BS}-\ac{RIS}-\ac{UE}. 
Since \ac{RIS} reflects the signal and focuses the energy into a specific direction, \ac{UE}'s location should be considered in the estimation process. However, in \cite{3,4,6,9,Alouini2,8,5,7,11,estimation-mm,estimation-mm2,compressed-sensing1,compressed-sensing2}, the location is always ignored and only scenarios with stationary \ac{BS}, \ac{RIS} and \ac{UE} are considered. Furthermore, it is proven in \cite{13,12} that the path loss is a function of reflection coefficients of \ac{RIS} which is always ignored in the channel estimation process when the phases are optimized for channel estimation.

In this article, we develop a general three-stage framework for the \ac{RIS}-aided communication network, where practical issues are considered in a realistic scenario. We summarize the main contributions of this paper as follows
\begin{itemize}
    \item First, we derive the relation between \ac{RIS} and \ac{MIMO} system by providing an accurate configuration for the \ac{RIS} reflection coefficients array, noting that the resultant array is equivalent to the steering vector of the \ac{MIMO} system that has \ac{UPA} in its antenna structure. The derived model for the \ac{RIS} is obtained from the free space far-field path loss model that is introduced in \cite{12}.
    \item Additionally, we optimize the reflected signal in any specific direction simply by controlling the phases, where the effect of channel \ac{BS}-\ac{RIS} is eliminated at the \ac{BS} side, and the \ac{UE} has the responsibility of estimating and compensating channel \ac{RIS}-\ac{UE}. For the first time, this optimization proposes one channel control (\ac{BS}-\ac{RIS}) for the \ac{RIS} design instead of endeavoring the total cascaded channel control.
    \item Next, a novel channel estimation scheme for mmWave \ac{RIS}-\ac{MIMO} system is proposed. This scheme is able to estimate both \ac{BS}-\ac{RIS} and \ac{RIS}-\ac{UE} channels separately, even though all \ac{RIS} elements are passive. Starting with estimating the \ac{BS}-\ac{RIS} channel $\mathbf{G}$ using hierarchical beam searching algorithm. Then, the \ac{RIS}-\ac{UE} channel $\mathbf{H}$ is estimated by adopting the iterative reweight algorithm that is introduced in \cite{Super-resolution} to estimate the channel path coefficients only, exploiting the resultant angles from the beam searching algorithm.
    \item Then, the proposed scheme enables \ac{RIS}-assisted communication to track mobile users. To the best of our knowledge, this has never been addressed in the literature and it is considered one of the most challenging tasks to be implemented by the \ac{RIS}. The parameters of channel $\mathbf{H}$ are tracked using well-known algorithms such as the \ac{EKF} algorithm.
   \item Finally, the mmWave \ac{RIS}-\ac{MIMO} framework is studied under practical and implementable assumptions. The proposed design of the \ac{RIS} reflection coefficients has low computational complexity and applicable in real-time scenarios. Meanwhile, all channel effects are considered in the design including path loss, fading, user's location, incident and reflected angles. We analytically show that our proposed design achieves better performance compared to the conventional methods under the same assumptions.
\end{itemize}  

The rest of this paper is organized as follows. Section \ref{section:sec2} discusses some assumptions available in the literature. Section \ref{section:sec3} depicts the system model of the proposed \ac{RIS} framework, and Section \ref{section:sec4} discusses how to control the \ac{RIS}'s reflection coefficients to realize beamforming/steering functionalities. The novel channel estimation scheme is introduced is Section \ref{section:sec5} followed by channel tracking approaches. In Section \ref{section:sec6}, the performance analysis is carried out, and Section \ref{section:sec7} concludes the paper. 

\textit{Notation:} bold uppercase $\mathbf{A}$, bold lowercase $\mathbf{a}$, and unbold letters $A,a$ are used to denote matrices, column vectors, and scale values, respectively. $|a|$ and $\angle a$ are the magnitude and phase of a complex number. $\|\mathbf{a}\|_F$, $\|\mathbf{a}\|_0$, and $\|\mathbf{a}\|_2$ are the Frobenius norm, $\ell_0$ pseudo-norm, and the $\ell_2$ norm. $(\cdot)^H$, $(\cdot)^T$, and $(\cdot)^{-1}$ denote the Hermitian transpose, transpose, and inverse. $\operatorname{diag}(\mathbf{a})$ is the diagonal matrix with the vector $\mathbf{a}$ on its diagonal. $\mathbb{C}^{{M\times N}}$ denotes the space of $M\times N$ complex-valued matrices and $\operatorname{vec}(\mathbf{A})$ is vectorizing the matrix $\mathbf{A}$. $A \otimes  B$ is the Kronecker product of $A$ and $B$ and symbol $j$ represents the imaginary unit of complex numbers with $j^2=-1$.

\section{Investigation about \ac{RIS}} \label{section:sec2}
In this section, three major concerns are discussed regarding path loss, channel model, and user tracking in \ac{RIS}-assisted networks. 

\subsection{If multipath, path loss, and beamforming are to be optimized at the same time, what should be the reflection coefficients of the \ac{RIS}?}
Prior works have investigated \ac{RIS} phases matrix based on different criteria. 
Some authors \cite{12,13,Marco-di-ranzo} specified the \ac{RIS} matrix $\boldsymbol{\phi}$ so that the path loss is minimized, while others \cite{2,6,7,8} used these phases to align all multipath and get rid of channel effects. And some other authors designed $\boldsymbol{\phi}$  for decreasing the error in channel estimation. Eventually, using the same parameters for multiple purposes at the same time will create a conflict, and in this case, one general multi-goal design of the phases is required.

\begin{figure}
    \centering
    \includegraphics[scale=0.23]{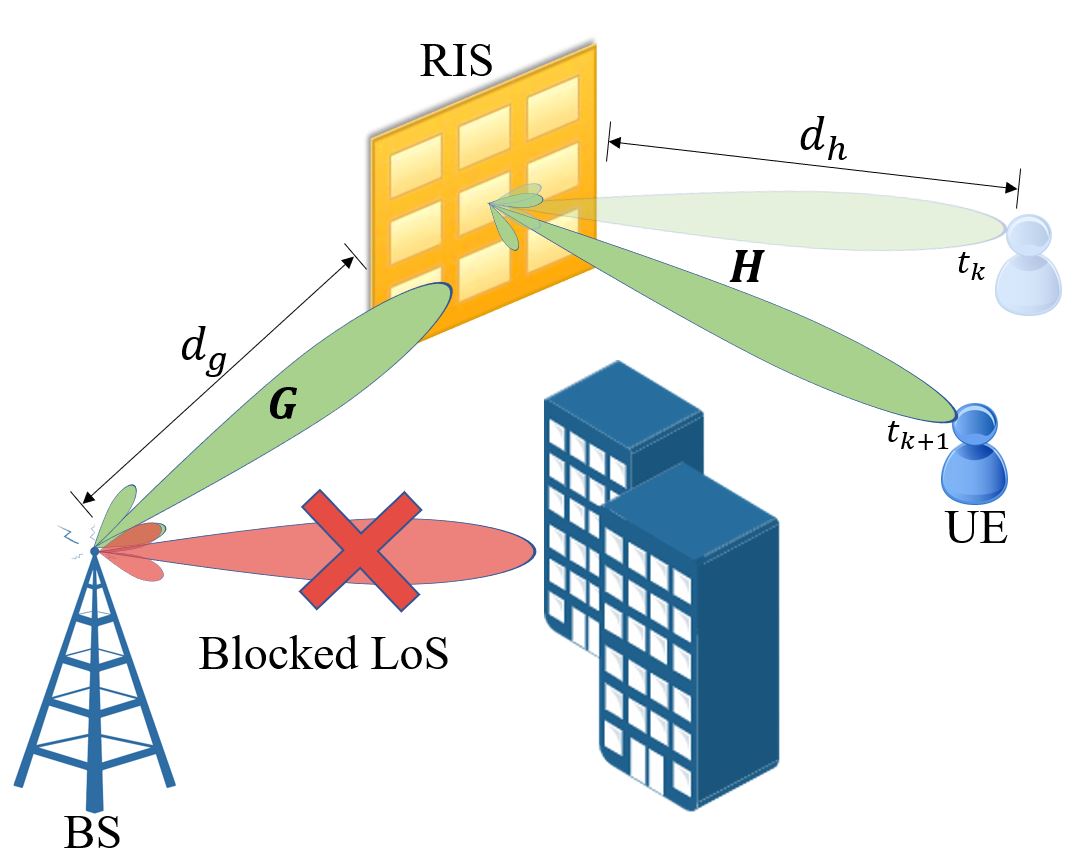} 
    \caption{RIS-aided communication system model.}
    \label{fig:systemmodel}
\end{figure}

Assuming that the \ac{RIS} elements are placed in a uniform rectangular shape, then the reflection coefficients of these elements is reflected by
\begin{equation}
    \boldsymbol{\phi} = \left[ \begin{array}{cccc}
    \phi_{1,1} & \phi_{1,2} & \cdots& \phi_{1,N_{RIS}}  \\
         \vdots& \ddots& & \vdots\\
    \phi_{N_{RIS},1} & \phi_{N_{RIS},2} & \cdots& \phi_{N_{RIS},N_{RIS}}\\ 
    \end{array} \right],
\end{equation}
where $\mathbf{\phi}_{n,m} =\gamma_{n,m}e^{j\alpha_{n,m}}$ is the $(n,m)$-th \ac{RIS} element's reflection coefficient, where $\alpha_{n,m}\in[0,2\pi)$ represents the phase shift induced by the $(n,m)$-th element in the \ac{RIS}, and $\gamma_{n,m}\in[0,1]$ stands for the reflection gain which will be considered unity throughout the paper i.e., $\gamma_{n,m}=1, \forall (n,m)$.
Another convenient representation of $\boldsymbol{\phi}$ in term of facilitating computations is defined as $\mathbf\Theta = \operatorname{diag}\{{\operatorname{vec}(\boldsymbol{\phi})}\}$.

Considering the system model that is shown in Fig. \ref{fig:systemmodel}, the reflected signals from each element of the \ac{RIS} are all aligned in phase to enhance the received signal power. It is assumed that the direction of the radiation is toward the center of the \ac{RIS} surface. For more simplification, let the dimensions of the each element be $d_x\times d_x$ and the total number is $M_{RIS}~=~N_{RIS}\times N_{RIS}$ elements. In this case, the free-space path loss is given as \cite{12}
\begin{equation}
\begin{aligned}
&\beta_{\mathrm{RIS}} = \frac{G_{t} G_{r} G N_{\mathrm{RIS}}^{4} d_{x}^{2} \lambda^{2} F\left(\theta_{t}, \varphi_{t}\right) F\left(\theta_{r}, \varphi_{r}\right) \gamma^{2}}{64 \pi^{3} d_{g}^{2} d_{h}^{2}} \\
& \times\bigg|\frac{\operatorname{sinc}\left(\frac{\pi N_{\mathrm{RIS}}}{\lambda}\left(\sin \theta_{t} \cos \varphi_{t}+\sin \theta_{r} \cos \varphi_{r}+\delta_{1}\right) d_{x}\right)}{\operatorname{sinc}\left(\frac{\pi}{\lambda}\left(\sin \theta_{t} \cos \varphi_{t}+\sin \theta_{r} \cos \varphi_{r}+\delta_{1}\right) d_{x}\right)} \\
& \frac{\operatorname{sinc}\left(\frac{\pi N_{\mathrm{RIS}} }{\lambda}\left(\sin \theta_{t} \sin \varphi_{t}+\sin \theta_{r} \sin \varphi_{r}+\delta_{2}\right) d_{x}\right)}{\operatorname{sinc}\left(\frac{\pi}{\lambda}\left(\sin \theta_{t} \sin \varphi_{t}+\sin \theta_{r} \sin \varphi_{r}+\delta_{2}\right) d_{x}\right)} \bigg|^{2},
\end{aligned}
\label{equ:path loss}
\end{equation}
where $\delta_{1}\left(m-\frac{1}{2}\right) d_{x}+\delta_{2}\left(n-\frac{1}{2}\right) d_{y}=\frac{\lambda \phi_{n, m}}{2 \pi}$ , $G_{t},G_{r}$  and ${G}$ are the gains of the transmitter antenna, receiver antenna, and the \ac{RIS}, respectively, $\lambda$ is the wavelength, $F(\theta,\varphi)$ is the normalized power radiation, $(\theta_t,\varphi_t)$ and $(\theta_r,\varphi_r)$ represent the elevation and the azimuth angles of the incidence wave and reflected wave, respectively, and $d_{g}$ and $d_{h}$ are the distance from the \ac{BS} to the \ac{RIS} and from the \ac{RIS} to the \ac{UE} respectively.
Let $(\theta_{des},\varphi_{des})$ be the angle from the \ac{RIS} to \ac{UE}, if $\theta_r= \theta_{des}$ and $\varphi_r=\varphi_{des}$, then (\ref{equ:path loss}) is maximized as 

\begin{equation}
\begin{aligned}
\beta_{max}(d,\theta,\varphi) &= \frac{G_{t} G_{r} G N_{\mathrm{RIS}}^{4} d_{x}^{2} \lambda^{2} F\left(\theta_{t}, \varphi_{t}\right) F\left(\theta_{r}, \varphi_{r}\right) \gamma^{2}}{64 \pi^{3} d_{g}^{2} d_{h}^{2}}.
\end{aligned}
\label{equ:path lossmax}
\end{equation}
This corresponds to the phases of the \ac{RIS} $\mathbf{\phi}_{n,m}$ being designed as follows
\begin{equation}
\begin{aligned}
    \angle{{\phi} _{n,m}^1}=\operatorname{mod} \left( -\dfrac {2\pi }{\lambda }d_x \bigg[ \left( \sin \theta _{t}\cos \varphi _{t}+\sin \theta_{des}\cos \varphi_{des}\right)\right. \\
   \left. ( m-\dfrac {1}{2}) + \left( \sin \theta _{t}\sin \varphi _{t}+\sin \theta _{des}\sin \varphi _{des}\right) ( n-\dfrac {1}{2} )\bigg],2\pi\right).
\end{aligned}
 \label{equ:phase3}
\end{equation}

\begin{figure*}[!t]
    \begin{center}
    \subfloat[$\mathbf{G}$ is an ideal channel ($\mathbf{G}=1$).]{\label{convPerf:1}\includegraphics[scale=0.42]{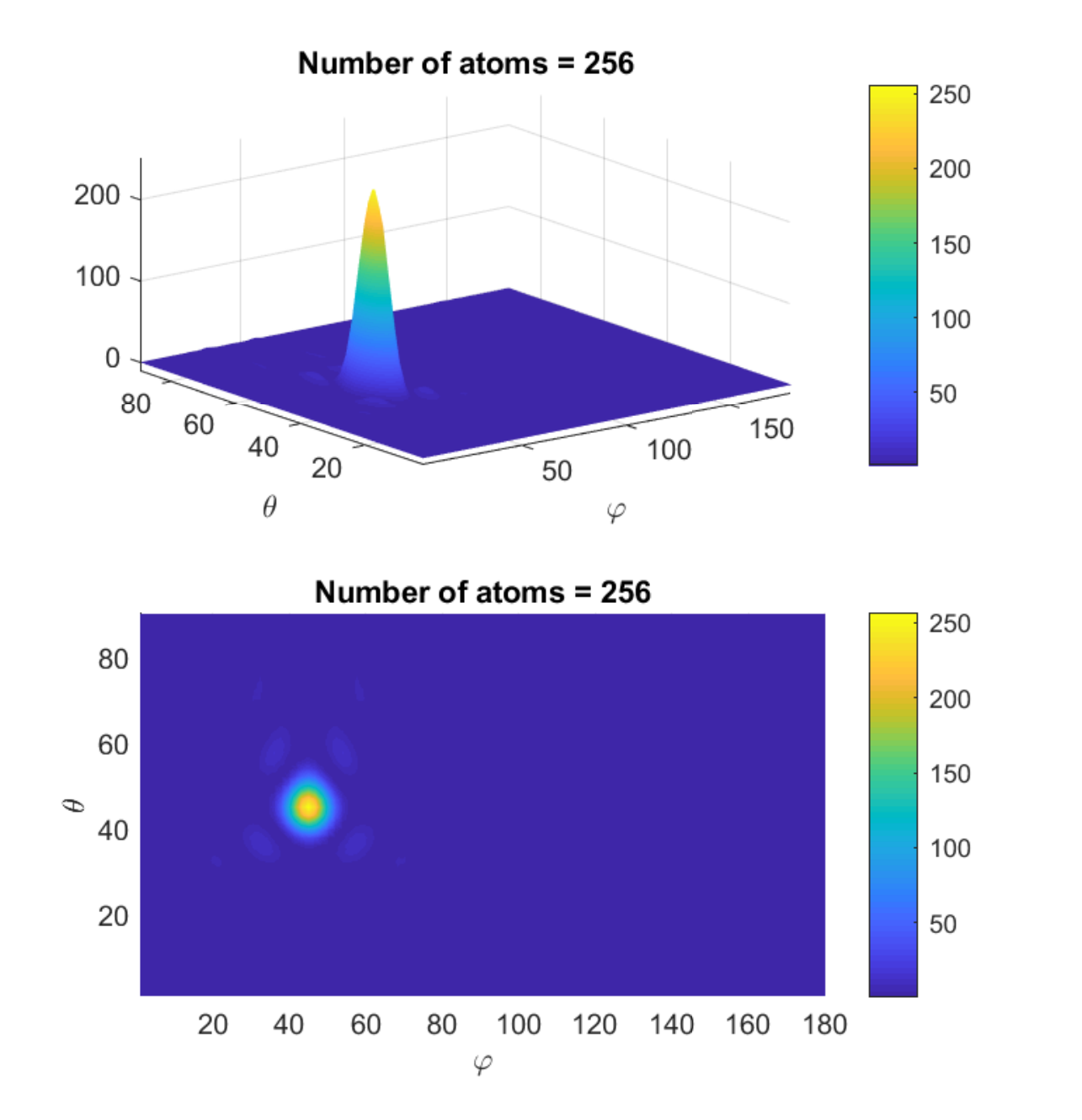}}
    \subfloat[$\mathbf{G}$ is a sparse channel with LoS path is the dominant.]{\label{convPerf:2}\includegraphics[scale=0.42]{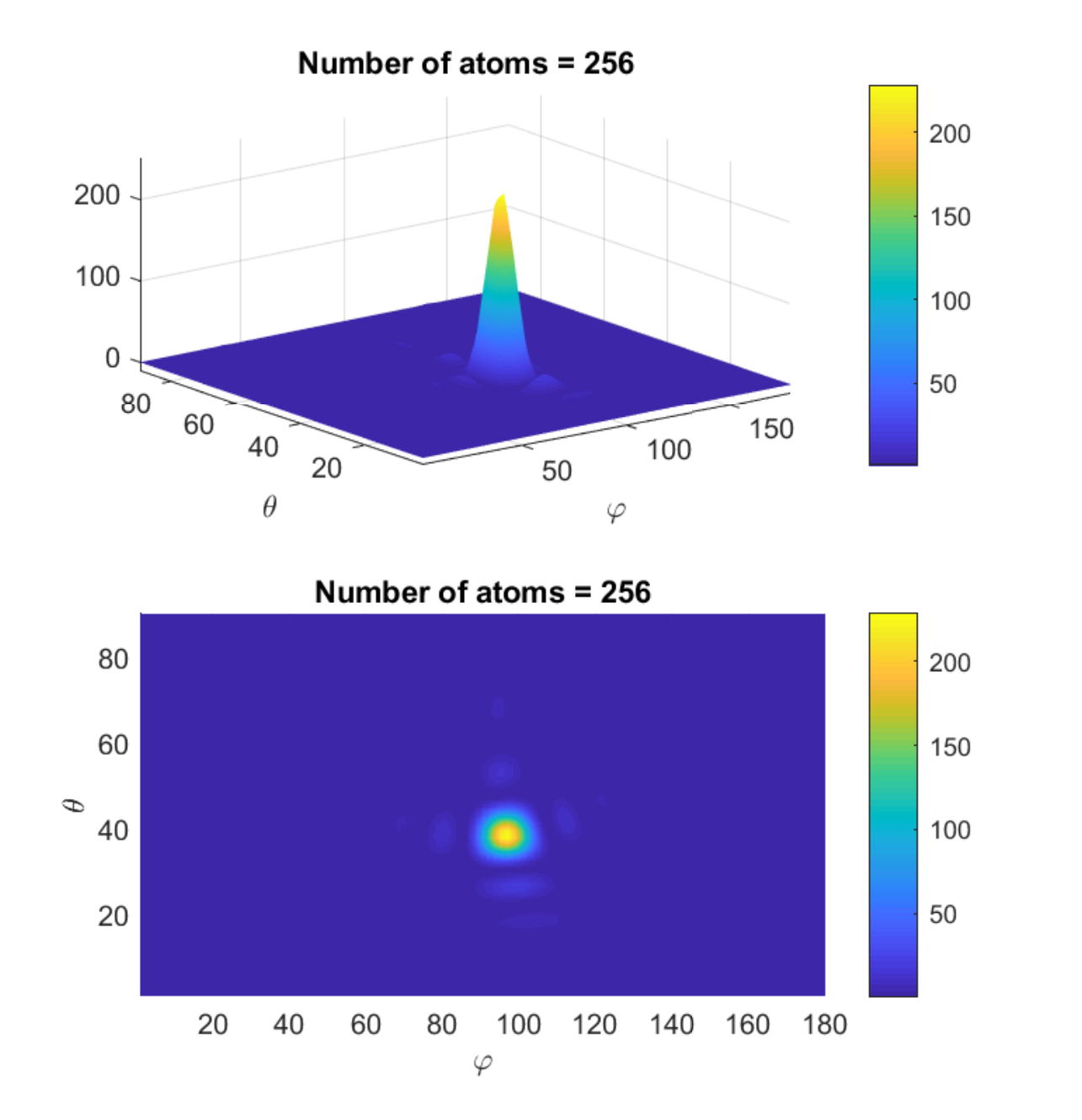}}
    \subfloat[$\mathbf{G}$ is a rich scattering channel.]{\label{convPerf:3}\includegraphics[scale=0.42]{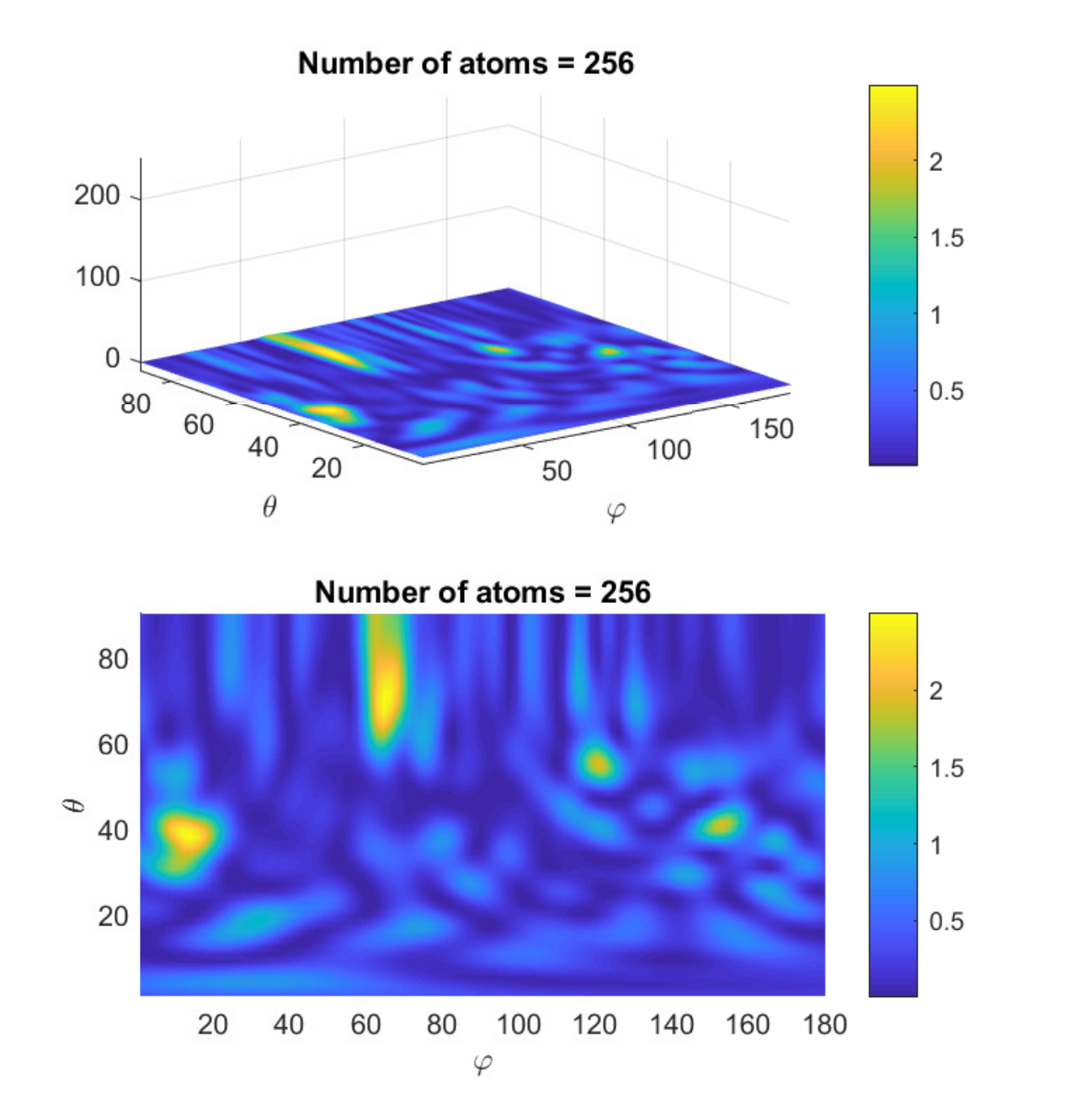}} 
    \\
    \end{center}
    \centering
    \caption{The power pattern of the reflected beam from the \ac{RIS} toward \ac{UE} under three conditions of channel $\mathbf{G}$}
    \label{fig:RISpower}
\end{figure*}

Fig. \ref{fig:RISpower}a shows the radiation pattern of the reflected signal when $\boldsymbol{\phi}$ is set to reflect the received signal in a specific direction $(\theta_{des},\varphi_{des})=(45^o,60^o)$, the power is focused in the desired direction and diminishes elsewhere.  
 
\ac{RIS} phases are optimized to obtain strongest channel impulse response
\begin{equation}
    \mathbf{\phi}_{n,m}^2 = \operatorname{max}_{\phi}\left(\left|\sum_{n=1}^{N}\sum_{m=1}^{M} {\phi _{n,m}} G_{n,m}H_{n,m} \right|^{2}\right),
\end{equation}
which implies $\mathbf{\phi}_{n,m}^2 = -\angle{\mathbf{G}_{n,m}\mathbf{H}_{n,m}}$. 
In the same paper \cite{6} however,  in channel estimation, $\boldsymbol{\phi}$ is chosen so that channel estimation error is minimized, for instance, the authors in \cite{5} claim that for minimizing \ac{MSE}, the reflection pattern of the \ac{RIS} should be chosen to be $N_{RIS}\times N_{RIS}$ \ac{DFT} matrix ${\phi}_{n,m}^3 = DFT^{N_{RIS}\times N_{RIS}}$. It is clearly seen that there is a contradiction of $\boldsymbol{\phi}^1 \neq\boldsymbol{\phi}^2 \neq \boldsymbol{\phi}^3$.

``\textit{In case of $\boldsymbol{\phi} = DFT^{N_{RIS}\times N_{RIS}}$, is it guaranteed that all multipath are aligned, and the signal is reflected toward the direction of user with maximum power, i.e., $P_r~\propto~M_{RIS}^2~\times~\beta_{max}$? If not, up to what extent the estimated channel is reliable? And if some power is received, what ensures that the source of this power is the \ac{RIS} and not other random reflectors?}"

Table \ref{table:cases} summarizes all the above-mentioned cases of different \ac{RIS}s' phases matrices compared to the ideal desired case. 
\begin{table} [t]
\begin{center}
\caption{Impact of $\phi$ design on the received power} 
\begin{tabular}{|c|c|}
 \hline
\textbf{Case}  & \textbf{Received signal's power $P_r$ } \\
 \hline
 Ideal/desired case  &  $P_r~\propto~\frac{M_{RIS}^2}{d_g^2~d_h^2}$     \\
\hline
$\boldsymbol{\phi}=\boldsymbol{\phi}^1$  & $P_r~\propto~\frac{A}{d_g^2~d_h^2},~ A\in [0,M_{RIS}^2] $ \\
\hline
$\boldsymbol{\phi}=\boldsymbol{\phi}^2$ & $P_r~\propto~B\frac{M_{RIS}^2}{d_g^2~d_h^2},~ B\in [0,1]$ \\
\hline
$\boldsymbol{\phi}=\boldsymbol{\phi}^3$ & $P_r~\propto~\frac{AB}{d_g^2~d_h^2}$ \\
\hline
Infinite reflector & $P_r~\propto~\frac{B}{(d_g+d_h)^2}$ \cite{goldsmith}\\
\hline
\end{tabular}
\label{table:cases}
\end{center}
\end{table} 
From this Table, it is concluded that for reliable channel estimation and communication using \ac{RIS}, the \ac{RIS}'s phases must be optimized taking into consideration beamforming/steering direction (\ac{UE}'s location) and the impact of channels $\mathbf{H}$ and $\mathbf{G}$. All the mentioned points here will be examined in detail in Section \ref{section:sec5}.

\subsection{What is the impact of channel $\mathbf{H}$ and $\mathbf{G}$ on the \ac{RIS} phases? Is it the same or different?} \label{section:sec2,2}

Using what was discussed in Section \ref{section:sec2}-A, and the basic properties of \ac{RIS} that are mentioned and proved in \cite{13,emilMyths}, let it is assumed a scenario where a signal is going to be reflected toward the same location as shown in Fig. \ref{fig:RISpower}a with the assumption that $\mathbf{G}$ is an ideal channel i.e., it has unitary gain. The predefined values of the phases would be given by (\ref{equ:path lossmax}), and the resulted radiation will be exactly as in Fig. \ref{fig:RISpower}a. However, when $\mathbf{G}$ is assumed to be a sparse channel where \ac{LoS} path between \ac{BS} and \ac{RIS} is the dominant path, it is observed that the beam is shifted toward a different direction than that of \ac{UE}'s location as shown in Fig. \ref{fig:RISpower}b. Also, when channel $\mathbf{G}$ is very rich scattering, the \ac{UE} will receive very low power from \ac{RIS} as shown in Fig. \ref{fig:RISpower}c, and in this case, the \ac{RIS} may react worse than a normal reflector (metallic surface, wall, etc.) \cite{3}. 
Therefore, for a \textbf{successful reflection}, $\mathbf{G}$ should be individually estimated and then equalized at the \ac{RIS} by simply reversing its effect.

After reflecting the beam, the \ac{UE} estimates and equalizes $\mathbf{H}$ to complete a \textbf{successful communication}. In a nutshell, \ac{RIS} performs two operations separately
\begin{enumerate}
    \item \textit{\textbf{Accumulation}}, where it collects all the energy received by each of its elements\footnote{This is the reason why the gain is proportional to $M_{RIS}^2$ \cite{13}} and then align them by cancelling the effects of channel $\mathbf{G}$.
    \item\textbf{\textit{Beamforming/Steering}}, the \ac{RIS} acts like a virtual \ac{BS}, and focuses or steers the incoming electromagnetic waves using (\ref{equ:phase3}) toward the \ac{UE}'s location \cite{Marco-di-ranzo}.   
\end{enumerate}

\subsection{Assuming that our system has mobility, which part of the cascaded channel is considered varying; $\mathbf{G}$, $\mathbf{H}$, or both? Can the \ac{UE} be tracked if the varying source is unknown?}
The majority of the state-of-art considers the user to be stationary, and \ac{BS} have always \ac{LoS} with \ac{RIS}. However, these assumptions are not realistic, and tend to limit the use of \ac{RIS}. Besides, they are the consequence of utilizing the cascaded channel model that is given by $\mathbf{H}_{\operatorname{cascaded}}\overset{\Delta}{=} \mathbf{G}\mathbf{H}$. This representation makes channel tracking in time almost impossible, since any change in $\mathbf{H}_{\operatorname{cascaded}}$ could be due to the change in $\mathbf{G}$, $\mathbf{H}$ or both. Also, in Section \ref{section:sec2}-B, it was shown that only $\mathbf{G}$ affects the phases of the \ac{RIS}, not $\mathbf{H}$. Furthermore, if $\mathbf{G}$ is estimated separately, estimating $\mathbf{H}$ is feasible and tracking \ac{UE}s becomes possible. Consequently, the assumption of having a mobile user is valid.
For instance, the authors in \cite{11} claim that it is almost impossible to estimate $\mathbf{G}$ via conventional channel estimation schemes, since the \ac{RIS} elements are passive. This claim will be disproved during this work.

\section{System model} \label{section:sec3}
Consider a narrowband mmWave \ac{MIMO} system, composed of one \ac{BS}, one \ac{RIS} and $k$ \ac{UE}s as depicted in Fig.\ref{fig:systemmodel}. Each, \ac{BS}, \ac{RIS} and \ac{UE}, are equipped with equidistant \ac{UPA}s as an antenna structure with half-wavelengthed distance between the antenna elements in which they have $M_{\operatorname{BS}}$, $M_{\operatorname{RIS}}$, and $M_{\operatorname{UE}}$ antenna elements, respectively. It is considered that the uplink and downlink transmissions are using a \ac{TDD} protocol that exploiting channel reciprocity for the CSI acquisition at the \ac{RIS} in both link directions. The \ac{BS} is assumed to have $M_{\operatorname{RF}}$ \ac{RF} chains where the number of these chains is much smaller than the antenna array elements and larger than the number of \ac{UE}s ($k\leq M_{\operatorname{RF}}<<M_{\operatorname{BS}}$) \cite{RF-chain}, while the \ac{UE} consists of only one \ac{RF} chain. The \ac{RIS} is placed near to the \ac{UE} side and far from the \ac{BS} to minimize the path loss effect \cite{12}. In order to fully utilize the functionality of the \ac{RIS}, the channel path between the \ac{BS} and \ac{UE} is assumed to be blocked by an obstacle. The \ac{RIS} structure is the same as the one proposed in Section \ref{section:sec2}-A. 

Assuming that $\boldsymbol{s}$ training symbols are transmitted via orthogonal precoding beams for each user, such that there is no inter-user interference. Under this assumption, we shall restrict the analysis to one representative \ac{UE} without loss of generality. Under the assumption of flat-fading and perfect timing and frequency synchronization \cite{MIMOchannelHeath}, the sparsity of the channel is exploited by using geometric channel modeling \cite{geometric-model-1,geometric-model-2}, giving that $\mathbf{G} \in\mathbb{C}^{M_{RIS} \times M_{\operatorname{BS}}}$ and $\mathbf{H}  \in\mathbb{C}^{M_{\operatorname{UE}}\times M_{RIS}}$ denote the channel between \ac{BS}-\ac{RIS} and \ac{RIS}-\ac{UE}, respectively. The $\mathbf{G} $ model is given as
\begin{equation}
\begin{array}{c}
    \mathbf{G}  = \sum_{l=1}^{L_{g}}z_{g,l} \mathbf{a}_{M_{RIS}}(\theta_{g,l}^{R},\varphi_{g,l}^{R})\mathbf{a}_{M_{BS}}^{H}(\theta_{g,l}^{B},\varphi_{g,l}^{B}),\\
   ~~~~~  = \mathbf{A}_{M_{RIS}}(\Omega_{R}) \operatorname{diag}(\mathbf{z_g})\mathbf{A}_{M_{BS}}^{H}(\Omega_{B}),
\end{array}
\end{equation}
where $L_{g}$ is the number of channel paths received at the \ac{RIS}, $\theta_{g,l}^{R},\varphi_{g,l}^{R}$ and $\theta_{g,l}^{B},\varphi_{g,l}^{B}$ are the elevation and azimuth  angles of \ac{AoA} and \ac{AoD} in each path, and $z_{g,l}$ is the complex channel coefficient between \ac{BS}-\ac{RIS} at $l$th path. $\mathbf{z_g}= [z_{g,1},z_{g,2}, ... ,z_{g,L_g}]^T$, $\Omega_{R} = [ (\theta_{g,1}^{R}, \varphi_{g,1}^{R}), (\theta_{g,2}^{R}, \varphi_{g,2}^{R}), ..., (\theta_{g,L_g}^{R}, \varphi_{g,L_g}^{R}) ]^T$, and $\Omega_{B} = [ (\theta_{g,1}^{B}, \varphi_{g,1}^{B}), (\theta_{g,2}^{B}, \varphi_{g,2}^{B}), ...,(\theta_{g,L_g}^{B}, \varphi_{g,L_g}^{B}) ]$. $\mathbf{a}_{M_{RIS}}$ is the array response vector of the \ac{UPA} \cite{UPA2} represented by 
\begin{equation}
  \mathbf{a}_{M_{i}}(\theta,\varphi) = \frac{1}{\sqrt{M_i}}\bigg(\mathbf{q}\big(\operatorname{sin}(\theta)\operatorname{cos}(\varphi)\big)\otimes\mathbf{p}\big(\operatorname{sin}(\theta)\operatorname{sin}(\varphi)\big)\bigg),
  \label{equ:steering}
\end{equation}
where $\mathbf{q}(u)=\left[1,e^{j\frac{2\pi d}{\lambda}u},...,e^{j\frac{2\pi d}{\lambda}(N_{x}-1)u} \right]^T$ and $\mathbf{p}(v)=\left[1,e^{j\frac{2\pi d}{\lambda}v},...,e^{j\frac{2\pi d}{\lambda}(N_{y}-1)v} \right]^T$, for $i \in  \{RIS, BS\}$.
Going through the same derivation, $\mathbf{H}$ is expressed as
\begin{equation}
\begin{array}{c}
    \mathbf{H}   = \sum_{l=1}^{L_{h}}z_{h,l} \mathbf{a}_{M_{UE}}(\theta_{h,l}^{U},\varphi_{h,l}^{U})\mathbf{a}_{M_{RIS}}^{H}(\theta_{h,l}^{R},\varphi_{h,l}^{R}),\\
~~~~~~  = \mathbf{A}_{M_{UE}}(\Psi_{U}) \operatorname{diag}(\mathbf{z_h})\mathbf{A}_{M_{RIS}}^{H}(\Psi_{R}),
\end{array}
\end{equation}
where $\Psi_{U} = [(\theta_{h,1}^{U},\varphi_{h,1}^{U}),(\theta_{h,2}^{U},\varphi_{h,2}^{U}),...,(\theta_{h,L_h}^{U},\varphi_{h,L_h}^{U})]^T$, $\Psi_{R} = [(\theta_{h,1}^{R},\varphi_{h,1}^{R}),(\theta_{h,2}^{R},\varphi_{h,2}^{R}),...,(\theta_{h,L_h}^{R},\varphi_{h,L_h}^{R})]$, and  $\mathbf{z_h}= [z_{h,1},z_{h,2},...,z_{h,L_h}]^T$. The overall channel $\mathbf{H}_{\operatorname{eff}}\in\mathbb{C}^{M_{UE}\times M_{BS}}$ between the \ac{BS}-\ac{RIS}-\ac{UE} is written as
\begin{equation}
    \mathbf{H}_{\operatorname{eff}} = \beta(d_g,d_h,\theta_{des},\varphi_{des}) \mathbf{H}   \mathbf\Theta \mathbf{G},
    \label{equ:Heffect}
\end{equation}
where $\beta(d_g,d_h,\theta_{des},\varphi_{des})$ is the total path loss given by (\ref{equ:path lossmax}).

\section{\ac{RIS} control} \label{section:sec4}
This section describes how to control the reflecting coefficients of the \ac{RIS} so that all manipulation (beamforming/steering) can occur in real-time. The authors in \cite{12} developed a path loss model proved analytically and experimentally. The optimal phases to beamform/steer a reflected signal in a predefined direction is represented in \eqref{equ:phase3}, and it can be expressed as   
\begin{equation}
\begin{aligned}
    \operatorname{vec}(\boldsymbol{\phi}) = \Lambda_x(\theta _{t},\varphi _{t},\theta_{des},\varphi _{des})\otimes\Lambda_y(\theta_{t},\varphi_{t},\theta_{des},\varphi_{des}),
\label{predefined_phases}    
\end{aligned}
\end{equation}
where $\Lambda_x(.)$ and $\Lambda_y(.)$ can be viewed as steering vectors on the elevation and the azimuth direction, respectively, with
\begin{equation}
\Lambda_x = [e^{j\frac{-N_{\operatorname{RIS}}}{2}\frac{2\pi}{\lambda}d_x\left( \Gamma_x\right)},\dots,e^{j\frac{N_{\operatorname{RIS}}}{2}\frac{2\pi}{\lambda}d_x\left( \Gamma_x\right)}]^T,
\end{equation}
and
\begin{equation}
\Lambda_y = [e^{j\frac{-N_{\operatorname{RIS}}}{2}\frac{2\pi}{\lambda}dy\left( \Gamma_y\right)},\dots,e^{j\frac{N_{\operatorname{RIS}}}{2}\frac{2\pi}{\lambda}dy\left( \Gamma_y\right)}]^T,
\end{equation}
where $\Gamma_x =  \sin \theta _{t}\cos \varphi _{t}+\sin \theta _{des}\cos \varphi _{des}$ and $\Gamma_y=\sin \theta _{t}\sin \varphi _{t}+\sin \theta _{des}\sin \varphi _{des}$.

For simplicity, we consider the case with maximum power accumulated i.e., $\theta _{t}\approx 0$ where the received beam is perpendicular to \ac{RIS} surface, then \eqref{predefined_phases} becomes a function of the destination angles only. 
\begin{equation}
\begin{aligned}
    \operatorname{vec}(\boldsymbol{\phi}) = \Lambda_x(\theta _{des},\varphi _{des})\otimes\Lambda_y(\theta _{des},\varphi _{des}).
\label{RIS_steering_vector}    
\end{aligned}
\end{equation}

As we can see, the reflection coefficients of \ac{RIS} are equivalent to the steering vector of a \ac{UPA} \cite{plannar-array} in the \ac{MIMO} model. To adjust the beam direction of the \ac{RIS}, we just need to multiply $\mathbf{\phi}$ with a weight vector which is designed to steer the \ac{RIS} elements array toward the desired direction. According to \eqref{RIS_steering_vector}, the array factor of an \ac{RIS} is given as
\begin{equation} 
\begin{aligned}
    AF_{\operatorname{RIS}} = \sum_{n= -N_{\operatorname{RIS}}/2}^{N_{\operatorname{RIS}}/2}\sum_{m= -N_{\operatorname{RIS}}/2}^{N_{\operatorname{RIS}}/2} w_{m,n}, 
\label{RIS_AF}    
\end{aligned}
\end{equation}
where $w_{m,n}=e^{j[m\frac{2\pi}{\lambda}d_x\left( \Gamma_x\right)+n\frac{2\pi}{\lambda}dy\left( \Gamma_y\right)]}$ denotes the weight vector.

\section{Proposed Channel Estimation} \label{section:sec5}
In order to apply the proposed estimation technique, the effective channel in (\ref{equ:Heffect}) can be rewritten as
\begin{equation}
    \mathbf{H}_{\operatorname{eff}} = \beta(d_g,d_h,\theta_{des},\varphi_{des}) \hat{\mathbf{H}}\mathbf\Theta \hat{\mathbf{G}}, 
\end{equation}
where $\hat{\mathbf{H}} = \mathbf{A}_{M_{UE}}(\Psi_{U}) \operatorname{diag}(\mathbf{z})\mathbf{A}_{M_{RIS}}^{H}(\Psi_{R})$ and $\hat{\mathbf{G}} =\mathbf{A}_{M_{RIS}}(\Omega_{R}) \operatorname{diag}(e^{j\angle{\mathbf{z_g}}})\mathbf{A}_{M_{BS}}^{H}(\Omega_{B})$, and $\mathbf{z}$ is the gain of the cascaded channel $\mathbf{G}$ and $\mathbf{H}$ is considered. Since $\mathbf{G}$ is directly responsible of altering the \ac{RIS} phases, it is more meaningful to represent it only in terms of $\angle{\mathbf{z_g}}$, and include the channel gain $|\mathbf{z_g}|$ into $\mathbf{H}$. Writing the channel in this form allows us to estimate $\hat{\mathbf{H}}$ and $\hat{\mathbf{G}}$ separately.

\subsection{Estimating \ac{BS}-\ac{RIS} channel}
Since mmWave channel is sparse and the new representation of channel \ac{BS}-\ac{RIS} has unit amplitude, the problem of estimating $\hat{\mathbf{G}} $ becomes equivalent to the estimation of ${(e^{j\angle{\mathbf{z_g}}})}$ of each path.
From Section \ref{section:sec2,2}, $\hat{\mathbf{G}}$ causes a shift in the reflected beam, and hence, estimating this shift leads to estimate $\hat{\mathbf{G}}$ itself.
This could be done in three steps. First, estimating \ac{AoA} and \ac{AoD} for the \ac{RIS} reflected signal. Next, substituting these angles in (\ref{RIS_AF}) to get the reflection coefficients of \ac{RIS} in the absence of $\hat{\mathbf{G}}$'s effect. Then, these coefficients are compared to the last coefficient set by the \ac{BS}, and subtracted from each other to get $\hat{\mathbf{G}}$.

\begin{figure}
    \centering
    \includegraphics[scale=0.24]{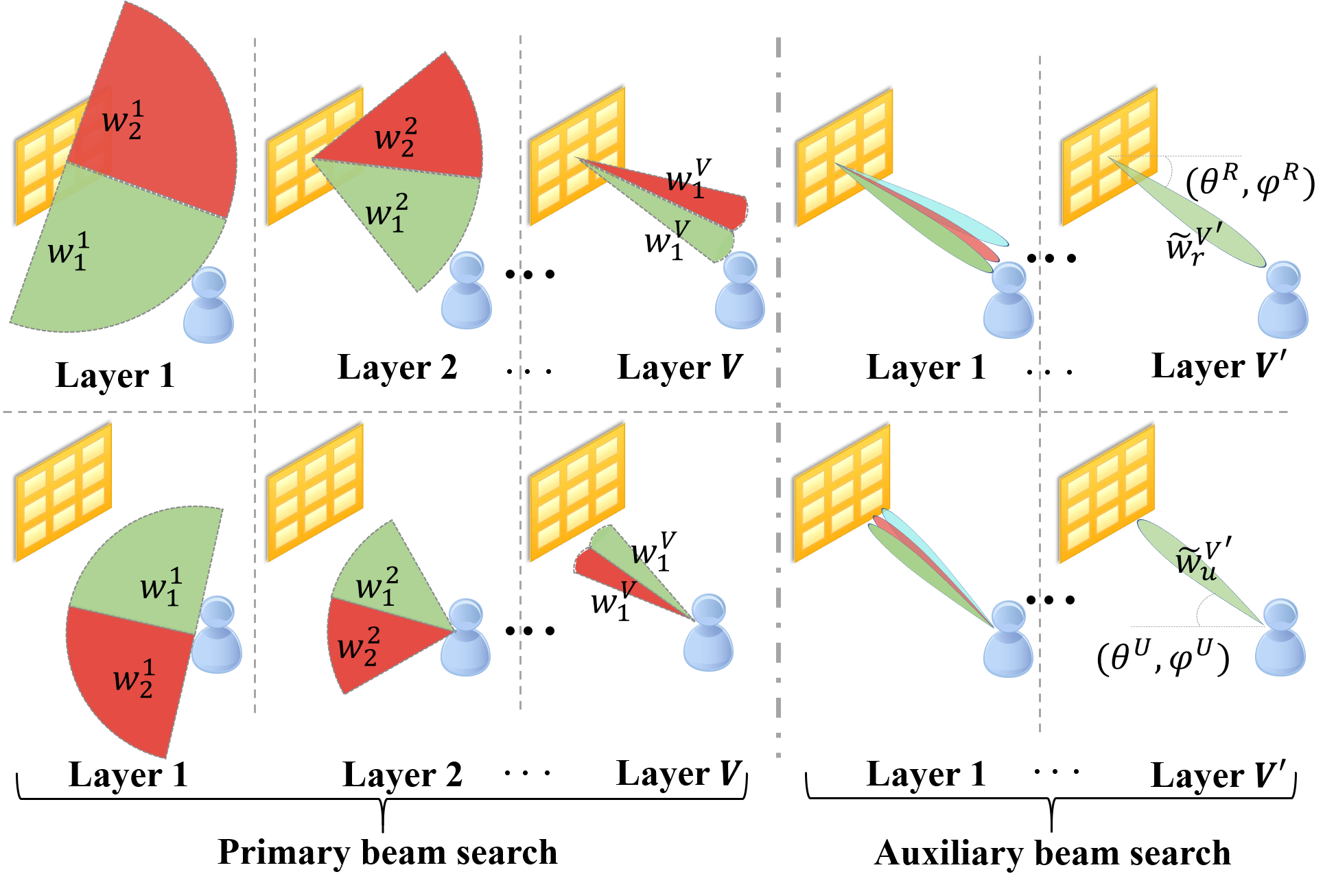} 
    \caption{Hierarchical beam searching algorithm procedures for channel estimation.}
    \label{fig:beamsearch}
\end{figure}

\subsubsection{Finding \ac{AoA} and \ac{AoD}} \label{section:sec5_A}
An exhaustive beam searching algorithm can be used in this case, where all possible angles are tested to find one optimal \ac{AoA}/\ac{AoD}. However, this approach requires a large amount of time due to its complexity \cite{Code-book-design2}. Thus, we adopt a two-stage beam training method depicted in Fig. \ref{fig:beamsearch} consisting of primary and secondary beam search \cite{Code-book-design}. For simplicity, the training procedure is described in azimuth only in the latter part of this subsection, and by using the same analogy the procedure in elevation can be deduced. 

The primary search will use hierarchical search to reduce the search time. As given in \cite{Code-book-design}, two-way tree stage search is used here at each layer. Let $w_n^l$ denotes the codeword of the $n^{th}$ beam vector at the $l^{th}$ layer, at each layer only $2^l$ antennas are activated. In total, there will be $\aleph$ possible beams and $V = \operatorname{log_2}(\aleph)$ layers, where each parent codeword $w_n^l$ has two child codewords $w_n^{l+1}$ and $w_{n+1}^{l+1}$. It is aimed to obtain $(\theta_{h,l}^{U},\varphi_{h,l}^{U})$ and $(\theta_{h,l}^{R},\varphi_{h,l}^{R})$ through multiple steps. Starting by testing four wide-beams in four successive time slots, where the \ac{RIS} uses $\mathbf{w}_r = [w_1^1,w_2^1]$ at reflecting mode\footnote{Note that by setting the phases of the \ac{RIS} according to equation \eqref{RIS_AF}, and setting the weighting vector to be any chosen codeword i.e., $ \mathbf{w} = (\operatorname{\mathbf{w}_r^{el}}\otimes\operatorname{\mathbf{w}_r^{az}})$, beam searching could be implemented at the \ac{RIS}.} and the \ac{UE} uses $\mathbf{w}_u = [w_1^1,w_2^1]$ at the receiving mode. The resulted signal from the $l^{th}$ stage can be written as 
\begin{equation}
    \boldsymbol{y}^{l}=\beta(d_g,d_h,\theta_{des},\varphi _{des})\mathbf{w}_u^H\hat{\mathbf{H}}   \mathbf{w}_r \mathbf{\hat{s}} +\mathbf{w}_u^H\boldsymbol{n},
\end{equation}
where $\mathbf{\hat{s}}=\hat{\mathbf{G}}\mathbf{s}$, $\mathbf{s}=[s_1, s_2, ..., s_Z]^T$ is $Z \times 1$ vector of transmitted symbols, and $\boldsymbol{n}$ is $Z \times 1$ complex Gaussian noise vector with zero-mean and variance $\sigma_o^2$. At each stage we search for the pair $(\mathbf{\tilde{w}}_r^l,\mathbf{\tilde{w}}_u^l)$ that satisfies the highest received \ac{SNR}, i.e.,
\begin{equation}
     \underset{\mathbf{w}_r,\mathbf{w}_u}{\operatorname{max}} \left(\left|\mathbf{w}_u^H\hat{\mathbf{H}}   \mathbf{w}_r \mathbf{\hat{s}}\right|^2\right)=\left|(\mathbf{\tilde{w}}_u^l)^H\hat{\mathbf{H}}   \mathbf{\tilde{w}}_r^l \mathbf{\hat{s}}\right|^2.
\end{equation}
After $V$ beam search, the optimum pair $(\mathbf{\tilde{w}}_r^V,\mathbf{\tilde{w}}_u^V)$ is obtained.
The elements of the primarily codebook matrix in azimuth of $K$ beam patterns, $ \tau$ discrete phase shift, and $N_{RIS}$ elements is given by \cite{Code-book-design}
\begin{equation}
    {w}^{az}_{n,k}= \operatorname{exp}\left(-j\frac{2\pi}{ \tau}\lfloor\frac{nk \tau}{K}\rfloor \right),
    \label{equ:waz}
\end{equation}
where $ n=0,1...N_{RIS}-1$ and $ k=0,1...K-1$. This codebook ensures that it has $\aleph$ possible states, and it fully spans the azimuth range. Similarly, the primary beam codebook matrix in elevation is given by
\begin{equation}
   {w}^{el}_{n,k}= \operatorname{exp}\left(-j\frac{2\pi}{ \tau}\lfloor\frac{nk \tau}{2K-2}\rfloor \right).
   \label{equ:wel}
\end{equation}

Second stage starts after acquiring the primary codebook, where we make a secondary beam search by rotating the primary beam to create higher-resolution secondary beams. These beams define the auxiliary codebook. Finally, $(\tilde{\mathbf{w}}_r,\tilde{\mathbf{w}}_u)$ is considered the optimum codebook.

Since the optimal transmission beam is represented by a weighting vector $\mathbf{w}=\operatorname{\tilde{\mathbf{w}}^{el}}\otimes\operatorname{\tilde{\mathbf{w}}^{az}}$, both \ac{AoA}/\ac{AoD} can be obtained. The \ac{AoA} from \ac{RIS} to \ac{UE} can be found as
\begin{equation}
(\theta_{h,l}^{U},\varphi_{h,l}^{U}) = \left(\operatorname{sin}^{-1} \big(\frac{-\lambda}{\tau}\lfloor\frac{k \tau}{K}\rfloor \big),\operatorname{sin}^{-1} \big(\frac{-\lambda}{\tau}\lfloor\frac{k \tau}{2K-2}\rfloor \big) \right).
\label{equ:AoAestimation}
\end{equation}

In our model, since the \ac{RIS} is located near to \ac{UE}, we assume that the antenna arrays of the \ac{UE} is always parallel to the \ac{RIS}, hence 
$(\theta_{h,l}^{R},\varphi_{h,l}^{R})=(\theta_{h,l}^{U},\varphi_{h,l}^{U})$ \cite{Tr4}.

\subsubsection{Estimating $\hat{\mathbf{G}}$}
If the \ac{RIS} phases are set to direct the beam of the reflected signal toward the \ac{UE}'s location $(\theta_{h,l}^{R},\varphi_{h,l}^{R})$, then the beam would be distorted and the radiation is shifted toward different direction due to the effect of channel $\hat{\mathbf{G}}$. Mathematically, this could expressed as
\begin{equation}
\mathbf{\hat{H}\Theta^{V'} \hat{G}}=\mathbf{\hat{H} \Theta}(\theta_{h,l}^{R},\varphi_{h,l}^{R}) \mathbf{G}_{opt} ,
\end{equation}
where $\mathbf{G}_{opt} = \mathbf{G}(\theta_{g,1}^{B},\varphi_{g,1}^{B},\theta_{g,1}^{R},\varphi_{g,1}^{R})$ \footnote{Note that the angles $\theta_{g,1}^{B},\varphi_{g,1}^{B},\theta_{g,1}^{R},\varphi_{g,1}^{R}$ are known from the fixed geometry of the deployment of \ac{RIS} and \ac{BS}.}, and $\mathbf{ \Theta^{V'} }$is the last configured set of phases by the \ac{BS} at the V'-th stage of beam searching process.
By exploiting the angles obtained from (\ref{equ:AoAestimation}) and by substituting them in \eqref{RIS_steering_vector}, $\hat{\mathbf{G}} $ can be estimated directly as
\begin{equation}
\begin{aligned}
\hat{\mathbf{G}}=\mathbf{(\Theta^{V'})^{-1}  \Theta}(\theta_{h,l}^{R},\varphi_{h,l}^{R}) \mathbf{G}_{opt}.
\end{aligned}
\label{equ:estimated_G}
\end{equation}
By adopting this design, we assure that the effect of $\hat{\mathbf{G}}$ is known and its effects are cancelled by the \ac{RIS}. Therefore, to set communication with any \ac{UE} at direction $(\theta_{des},\varphi_{des})$ throughout the \ac{RIS}, we simply set the phases by 
\begin{equation}
\mathbf{\Theta} =\mathbf{\Theta}(\theta_{des},\varphi_{des})\mathbf{G}_{opt}\hat{\mathbf{G}}^H(\hat{\mathbf{G}}\hat{\mathbf{G}}^H)^{-1}.
\label{equ:Theta_design}
\end{equation}
Please refer to Appendix. A\hfill$\blacksquare$

By substituting \eqref{equ:Theta_design} in the total channel we obtain 
\begin{equation}
\mathbf{\hat{H} \Theta \hat{G}}=\mathbf{\hat{H}}\mathbf{\Theta}(\theta_{des},\varphi_{des}) \mathbf{G}_{opt},
\end{equation}
where $\mathbf{\Theta}(\theta_{des},\varphi_{des})$ is set for any desired location, and the channel estimation problem is reduced to estimate $\mathbf{\hat{H}}$ only which will be explained in the next subsection. 

\subsection{Estimating \ac{RIS}-\ac{UE} channel}
Without loss of generality, assuming one \ac{RF} chain is activated at the \ac{BS} side and $Z$ symbols are transmitted, channel estimation model given in \cite{Super-resolution} is adopted here to estimate path gains of all paths. For that, the system model is given as 
\begin{equation}
\operatorname{\boldsymbol{y}}=\boldsymbol{Q}^H\mathbf{H}_{\operatorname{eff}}\boldsymbol{F}\mathbf{s} + \boldsymbol{Q}^H \boldsymbol{n},
\end{equation}
where $\boldsymbol{y}\in\mathbb{C}^{Z \times 1}$ is the received signal at \ac{UE}, and $\boldsymbol{Q}\in\mathbb{C}^{M_{UE}\times Z}$ and $\boldsymbol{F}\in\mathbb{C}^{M_{BS}\times Z}$ are the hybrid combining and the precoder matrices, respectively. 
The received signal at the \ac{UE} can be explicitly expressed as
\begin{equation}
\operatorname{\boldsymbol{y}}=\beta(d_g,d_h,\theta_{des},\varphi_{des})\boldsymbol{Q}^H \hat{\mathbf{H}}   \mathbf\Theta \hat{\mathbf{G}} \boldsymbol{F}\mathbf{s} + \boldsymbol{Q}^H \boldsymbol{n}.
\label{equ:receivesignal}
\end{equation}

Assuming $\mathbf{x}=\mathbf\Theta \hat{\mathbf{G}} \boldsymbol{F}\mathbf{s} \in\mathbb{C}^{{M_{RIS}\times 1}}$, where each element $\mathbf{x}_i$ is the transmitted symbol. For channel estimation, we will transmit known symbols at known indices, each received signal corresponding to a transmitted pilot symbol at $u$ time slot is given as  
\begin{equation}
    y_{p,u}=\beta(d_g,d_h,\theta _{des},\varphi _{des})\boldsymbol{q}^{H}_{u} \hat{\mathbf{H}}  x_{p,u} + \boldsymbol{q}^{H}_{u} n_{p,u}.
\end{equation}
Within $U$ time slots, $U_p$ different pilot sequences are sent in each time slot, and  $\boldsymbol{y}_p=\beta(d_g,d_h,\theta _{des},\varphi _{des})\boldsymbol{Q}^H\hat{\mathbf{H}}  \mathbf{x}_p+ \boldsymbol{Q}^H \boldsymbol{n}_p$, where $\boldsymbol{y}_p = [{y}_{p,1},{y}_{p,2},...,{y}_{p,U}]^T$ and $\boldsymbol{Q} = [\boldsymbol{q}_{1},\boldsymbol{q}_{2},...,\boldsymbol{q}_{U}]^T$. By setting $\boldsymbol{Y} = [\boldsymbol{y_{1}},\boldsymbol{y_{2}},..,\boldsymbol{y_{p}},..,\boldsymbol{y}_{U_p}]^T$, $\mathbf{X} = [\mathbf{x}_1,\mathbf{x}_2,..,\mathbf{x}_{U_P}]^T$ and $\boldsymbol{N} = [\boldsymbol{n}_1,\boldsymbol{n}_2,..,\boldsymbol{n}_{U_p}]^T$, we get
\begin{equation}
    \boldsymbol{Y} = \boldsymbol{Q}^H\hat{\mathbf{H}} \mathbf{X}+ \boldsymbol{Q}^H\boldsymbol{N}.
\end{equation}
Using the fact that the mmWave channel is sparse, the estimation of the channel $\hat{\mathbf{H}}$ become equivalent to the estimation of $\mathbf{z}$, $\Psi_{U}$ and $\Psi_{R}$, and the problem is formulated as
\begin{equation}
 \underset{\mathbf{z}, \Psi_{U}, \Psi_{R}}{\operatorname{min}} P_1(\mathbf{z}, \Psi_{U}, \Psi_{R}) \triangleq  \|\hat{\mathbf{z}}\|_0, s.t. \|\boldsymbol{Y}-\boldsymbol{Q}^H\Tilde{\mathbf{H}}\mathbf{X}\|_F \leqslant \epsilon,
\end{equation}
where $ \|\hat{\mathbf{z}}\|_0$ represents the number of non-zero elements, i.e., the sparsest solution of the sparse channel $\Tilde{\mathbf{H}}$, $\Tilde{\mathbf{H}}$ is the estimated channel matrix for $\hat{\mathbf{H}}$, and $\epsilon$ is the estimation error tolerance.

Since the log-sum penalty is more sparsity encouraging, the log-norm instead of $\|\hat{\mathbf{z}}\|_0$ can be used here \cite{regularization}. In addition, both $\Psi_{U}, \Psi_{R}$ are already obtained in Section V-A using the beam searching algorithm, thus the optimization is performed according to $\mathbf{z}$ only, and the problem $P_1$ is given as
\begin{equation}
\underset{\mathbf{z}}{\operatorname{min}}\quad P_2(\mathbf{z}) \triangleq \sum_{l=1}^{L_h}\log( |\hat{\mathbf{z}}|^2+\delta), s.t, \|\boldsymbol{Y}-\boldsymbol{Q}^H\Tilde{\mathbf{H}}\mathbf{X}\|_F \leqslant \epsilon,
\end{equation}
where $\delta$ ensures that the logarithmic function is always in its domain of definition. In addition to minimizing the number of paths, minimizing the channel estimation error is needed. Hence, a regularization parameter $\zeta>0$ is added, and $P_2$ is reshaped to the following optimization problem  
\begin{equation}
\underset{\mathbf{z}}{\operatorname{min}}\quad P_3(\mathbf{z}) \triangleq \sum_{l=1}^{L_h}\log( |\hat{{z}}|^2+\delta)+\zeta \|\boldsymbol{Y}-\boldsymbol{Q}^H\Tilde{\mathbf{H}}\mathbf{X}\|_F^2.
\end{equation}
It turned out that the minimization of $P_3$ is equivalent to the minimization of the iterative surrogate function \cite{regularization}
\begin{equation}
\underset{\mathbf{z}}{\operatorname{min}}\quad P_4^{(i)}(\mathbf{z}) \triangleq \zeta^{-1}\mathbf{z}^H\mathbf{D}^{(i)}\mathbf{z}+ \|\boldsymbol{Y}-\boldsymbol{Q}^H\Tilde{\mathbf{H}}\mathbf{X}\|_F^2,
\label{equ:minP4}
\end{equation}
where $\mathbf{D}^{(i)}$ is expressed as
\begin{equation}
\mathbf{D}^{(i)} = \operatorname{diag}\bigg( \frac{1}{|\hat{{z}}_1^{(i)}|^2+\delta} \frac{1}{|\hat{{z}}_2^{(i)}|^2+\delta} \cdots  \frac{1}{|\hat{{z}}_{L_h}^{(i)}|^2+\delta} \bigg),
\end{equation}
and $\hat{\mathbf{z}}^{(i)}$ is the estimate of $\mathbf{z}$ at the $i$th iteration. 
Then, the optimization of (\ref{equ:minP4}) becomes as follows
\begin{equation}
 P_4^{(i)}(\mathbf{z})= \zeta^{-1}\mathbf{z}^H\mathbf{D}^{(i)}\mathbf{z}+ \sum_{p=1}^{U_p}\|\boldsymbol{y}_{p}-\mathbf{T}_{p}\mathbf{z}\|_2^2,
\end{equation}
where $\mathbf{T_p}=\boldsymbol{Q}^H\mathbf{A}_{M_{UE}}(\Psi_{U}) \mathbf{A}_{M_{RIS}}^{H}(\Psi_{R})\mathbf{x}_p$.
\begin{equation}
\begin{aligned}
  P_4^{(i)}(\mathbf{z})= \zeta^{-1}\mathbf{z}^H\mathbf{D}^{(i)}\mathbf{z}+ \sum_{p=1}^{U_p}(\boldsymbol{y}_p-\mathbf{T}_p\mathbf{z})^H(\boldsymbol{y}_p-\mathbf{T}_p\mathbf{z}) \\
  = \mathbf{z}^H\bigg(\zeta^{-1}\mathbf{D}^{(i)}+ \sum_{p=1}^{U_p}\mathbf{T}_p^H\mathbf{T}_p\bigg)\mathbf{z}-\mathbf{z}^H\bigg(\sum_{p=1}^{U_p}\mathbf{T}_p^H\boldsymbol{y}_p\bigg)\\
  -\bigg(\sum_{p=1}^{U_p}\boldsymbol{y}_p^H\mathbf{T}_p\bigg)\mathbf{z}+\bigg(\sum_{p=1}^{U_p}\boldsymbol{y}_p^H\boldsymbol{y}_p\bigg).
\end{aligned}
\label{equ:Prob4}
\end{equation}
For optimizing (\ref{equ:Prob4}), the next step is obtained
\begin{equation}
 \frac{\partial P_4^{(i)}(\mathbf{z})}{\partial\mathbf{z}}=\mathbf{z}^H\bigg(\zeta^{-1}\mathbf{D}^{(i)}+ \sum_{p=1}^{U_p}\mathbf{T}_p^H\mathbf{T}_p\bigg)-\bigg(\sum_{p=1}^{U_p}\boldsymbol{y}_p^H\mathbf{T}_p\bigg) = 0.
\end{equation}
Therefore, the optimal $\hat{\mathbf{z}}$ that corresponds to the best estimation of $\Tilde{\mathbf{H}}$ at the $i$th iteration is given by
\begin{equation}
\begin{aligned}
  \mathbf{z}_{opt}^{(i)} \triangleq \bigg(\zeta^{-1}\mathbf{D}^{(i)}+ \sum_{p=1}^{U_p}\mathbf{T}_p^H\mathbf{T}_p\bigg)^{-1}\bigg(\sum_{p=1}^{U_p}\mathbf{T}_p^H\boldsymbol{y}_p\bigg).\\
  \triangleq\bigg(\zeta^{-1}\mathbf{D}^{(i)}+ \sum_{p=1}^{U_p}\mathbf{T}_p^H\mathbf{T}_p\bigg)\bigg(\sum_{p=1}^{U_p}\boldsymbol{y}_p^H\mathbf{T}_p\bigg)^{-1}.
\end{aligned}
\label{Z_opt}
\end{equation}

In this iterative method, $\zeta$ is designed to be adaptive to fit both a sparser estimation and a fast search. It is investigated in details in \cite{regularization,Super-resolution}.

\begin{figure}[h]
    \centering
    \includegraphics[width=8.6cm]{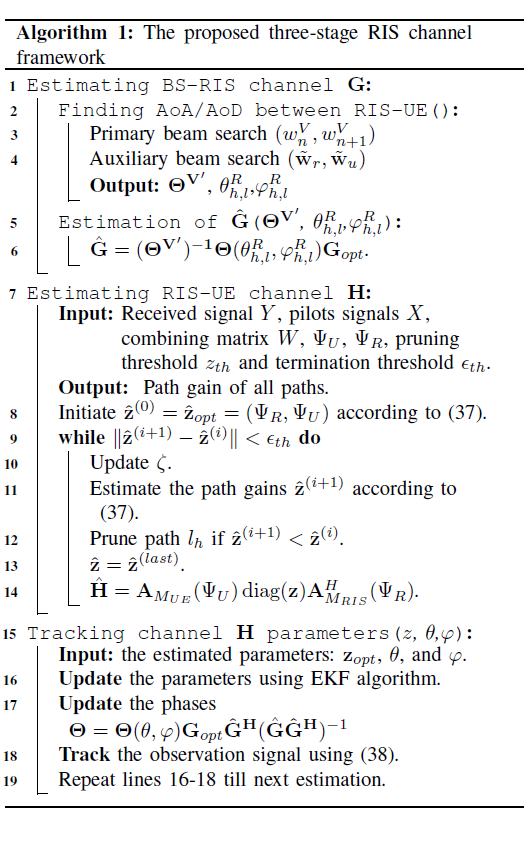}
\end{figure}

\begin{figure*}[h]
    \begin{center}
    \subfloat[]{\label{convPerf:Dectionary1}\includegraphics[scale=0.29]{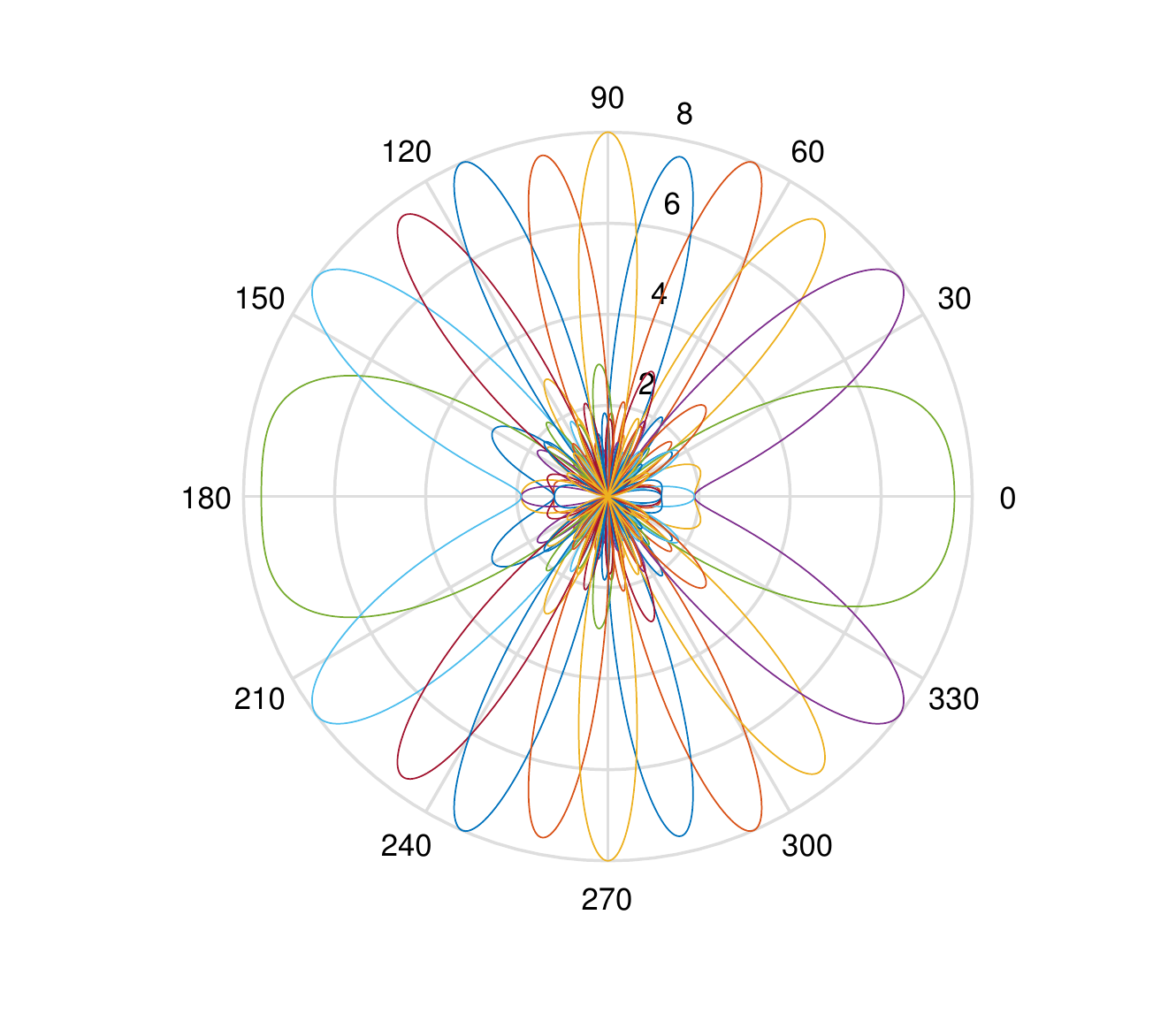}}
    \subfloat[]{\label{convPerf:Dectionary2}\includegraphics[scale=0.29]{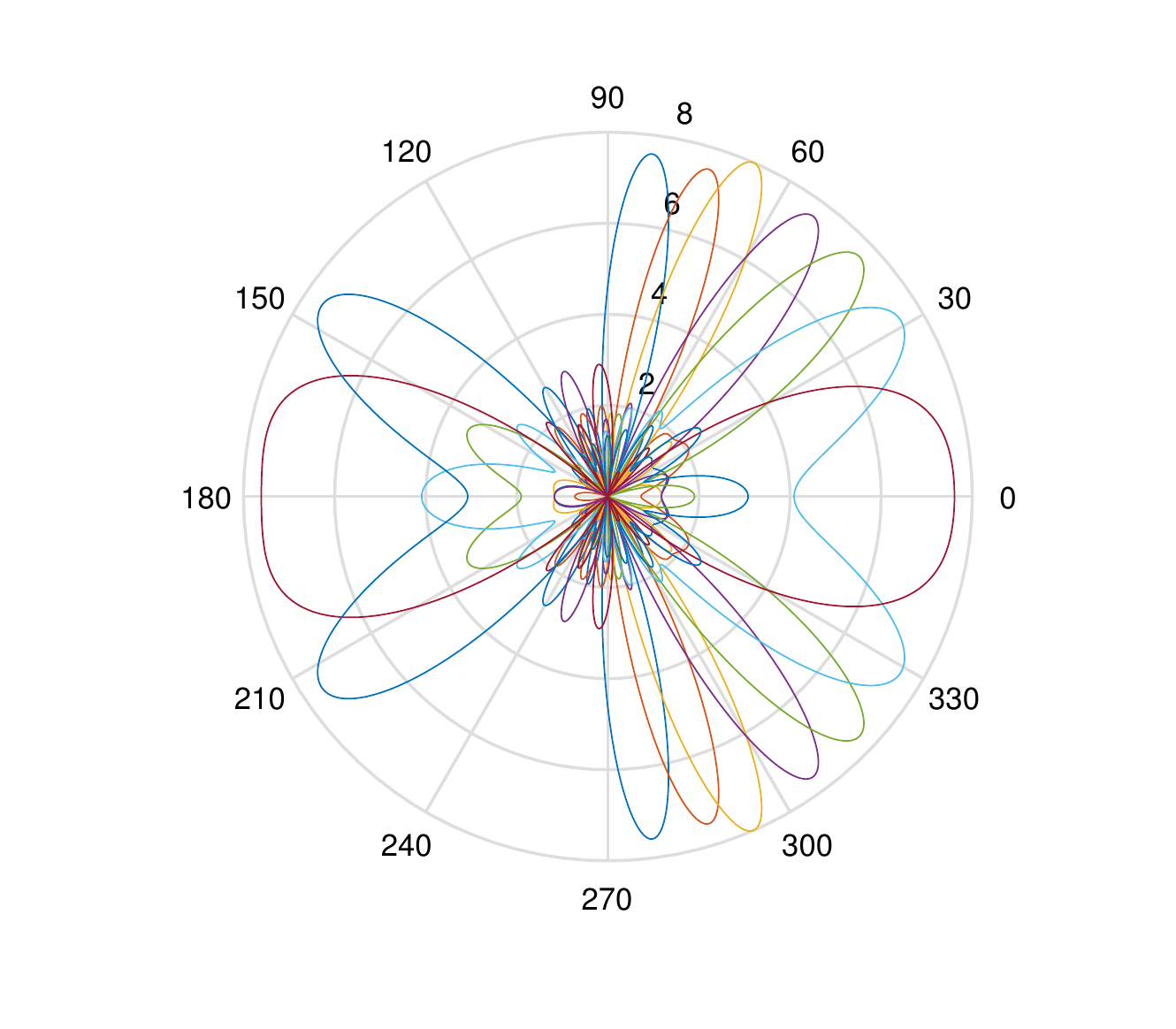}} 
    \subfloat[]{\label{convPerf:Dectionary3}\includegraphics[scale=0.29]{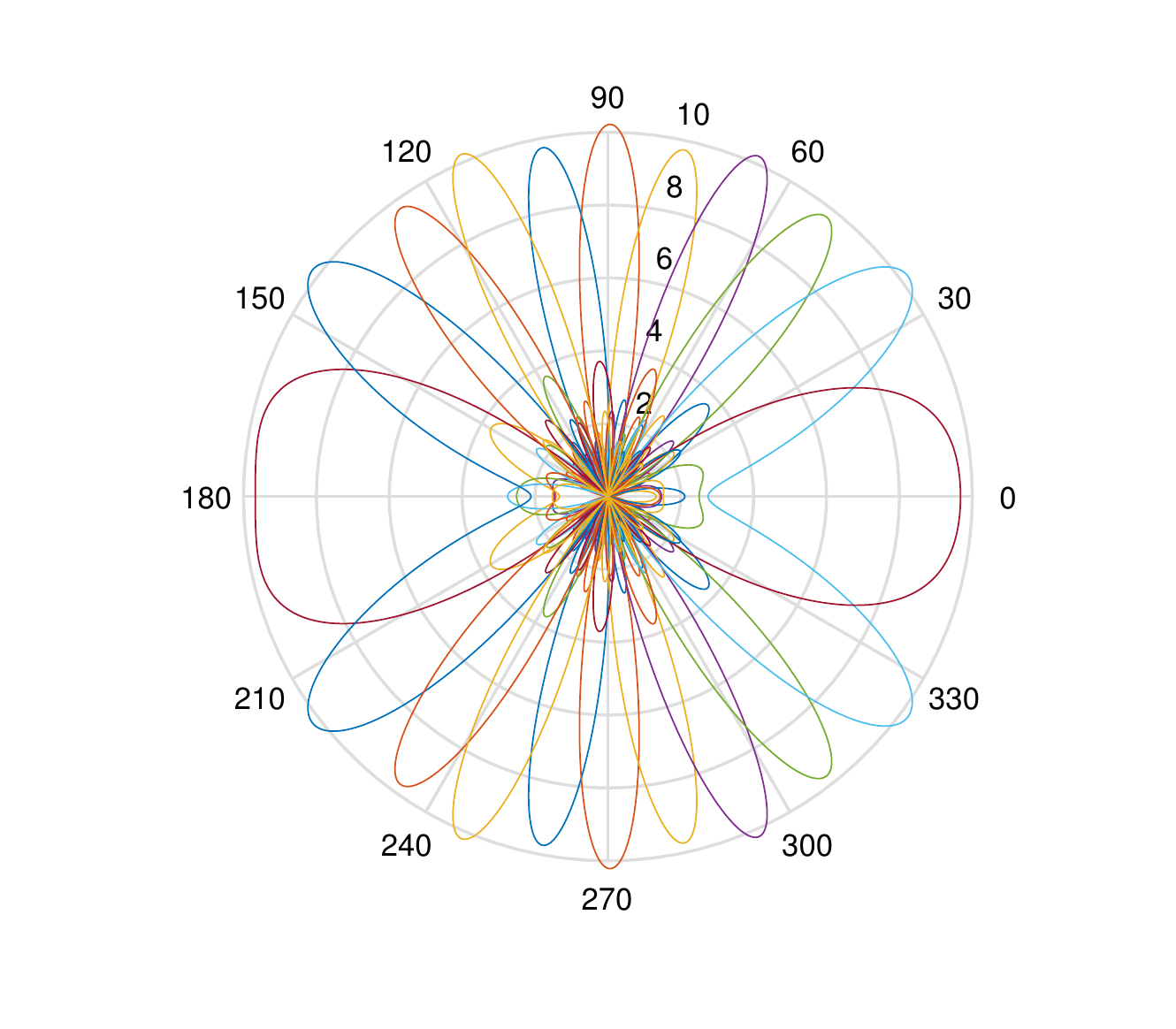}}  
    \subfloat[]{\label{convPerf:Dectionary4}\includegraphics[scale=0.29]{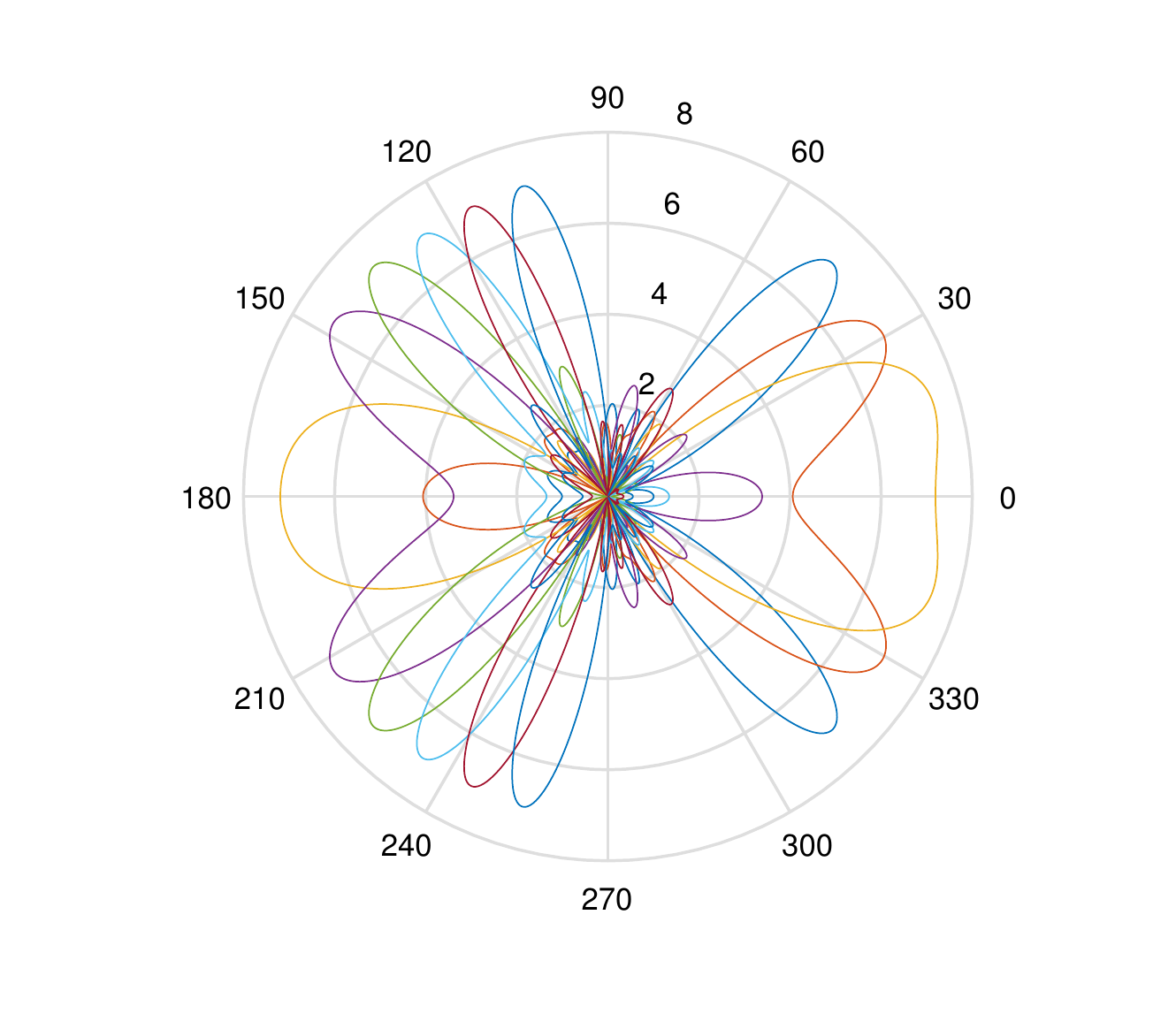}}  
    \\
    \end{center}
    \centering
    \caption{Primary beam patterns $N_{RIS}=8$, $\tau=5$, $K=10$ when channel $\mathbf{G}$ is ideal in (a) azimuth and (b) elevation domains, and when channel $\mathbf{G}$ is geometric model with $L_g=5$ paths in (c) azimuth and (d) elevation domains.}
    \label{fig:codebook}
\end{figure*}

\subsection{Channel Tracking} 
After estimating the channel parameters, i.e., channel coefficients, \ac{AoA}, and \ac{AoD}, and since the \ac{UE} is under mobility assumption, a channel tracking approach has been introduced here to avoid often channel estimation by tracking the channel parameters. The channel tracking algorithms are significantly fast, reliable, and robust which allow efficient data transfer between transmitters and receivers in mmWave communications. Channel tracking in mmWave systems is firstly presented in \cite{Tr1}, where an \ac{EKF} based tracking algorithm is proposed to track \ac{AoA}/\ac{AoD} while the channel coefficient remains constant. However, the method has difficulties to track in a fast-changing channel environment since it requires pre-requisites for a full scan that causes long time measurement. To decrease the measurement time and provide a more suitable tracking algorithm, the authors in \cite{Tr2} proposed an alternative solution that requires only a single measurement with \ac{EKF} estimation and a beam switching design. Additionally, \ac{LMS} and \ac{BiLMS} algorithms are introduced in \cite{Tr3} where advantages of both algorithms are presented compared to \ac{EKF} algorithm on imperfect \ac{CSI} conditions while having faster convergence characteristics as \ac{SNR} increases. However, both algorithms are not suitable for higher nonlinearity systems. Therefore, the \ac{EKF} tracking algorithm is used in our \ac{RIS}-assisted framework due to its low complexity and good tracking performance.

The tracking algorithm starts with setting a pair of transmitting and receive beams according to the estimated elevation and azimuth  \ac{AoA}/\ac{AoD} from the channel estimator. One main point that should be taken into consideration is that while tracking, the predicted channel parameters should stay close to the actual values so that the \ac{UE} stays within half of the beamwidth. Otherwise, if the tracking is no longer reliable or the path of the beams does not exist anymore, the channel parameters should be re-estimated.  

The discrete-time model for the received signal symbol period at \ac{UE} side is given in (\ref{equ:receivesignal}). Assuming that each vector in $\boldsymbol{F}$ is given by $\boldsymbol{f}=\mathbf{a}_{M_{BS}}(\theta,\varphi)$ for the \ac{LoS} path.
In order to start the tracking process, the measurement function should be known. From (\ref{equ:receivesignal}), the measurement function is used to track the observation signal and can be given as
\begin{equation}
    \boldsymbol{g}_{\operatorname{measure}} = \beta(d_g,d_h,\theta _{des},\varphi _{des})\boldsymbol{Q}^H \hat{\mathbf{H}}   \mathbf\Theta \hat{\mathbf{G}} \boldsymbol{F},
    \label{equ:measure}
\end{equation}
where $\boldsymbol{g}_{\operatorname{measure}}$ depends on the channel parameters including path coefficients, elevation and azimuth  \ac{AoD}/\ac{AoA} angles from both channels; \ac{BS}-\ac{RIS} and \ac{RIS}-\ac{UE}. \ac{EKF} algorithm \cite{Tr2} is used to track these parameters.
To evaluate the performance of the tracking process, a state evolution model is needed for the tracked parameters. A first-order Gaussian-Markov model is adopted for the path coefficient evolution over the time, while a Gaussian process noise model is assumed for the elevation and azimuth  \ac{AoD}/\ac{AoA} \cite{Tr2,Tr3,Tr4}. 

The proposed three-stage \ac{RIS} framework is summarized in Algorithm 1.

\section{Simulation results} \label{section:sec6}
In this section, simulation results are presented to evaluate the performance of the proposed \ac{RIS}-assisted framework. 
The simulation parameters are provided in Table \ref{table:simulationparameter}.

\begin{table}[h]
\caption{Simulation Configuration }
\centering
\begin{tabular}{l|l}
\hline
\multicolumn{1}{c|}{Parameters}                          &      \multicolumn{1}{c}{Value}   \\ \hline
Operating frequency $f_c $                               &  28 GHz         \\ \hline
Channel paths  $L_g$                                     &  5             \\ \hline
Channel paths $L_h$                                      &  1             \\ \hline
Antenna array size at \ac{BS} $M_{BS}$                   &  256              \\ \hline
Distance between antenna elements $d$                    &  $\lambda/2$      \\ \hline
Number of beam pattern in the codebook $K$               &  10             \\ \hline
Estimation error tolerance $\epsilon$                    &  1e-8              \\ \hline
\end{tabular}
\label{table:simulationparameter}
\end{table}

\subsection{Channel Estimation Performance}
At the first step of the proposed framework, we implement a two-stage beam search algorithm to determine the \ac{UE}'s location i.e., the \ac{AoD}/\ac{AoA} in elevation and azimuth  domains. Fig. \ref{fig:codebook} illustrates the primary beam patterns reflected from the \ac{RIS}, where $N_{RIS} = 8$ elements, each elements can perform $5$ discrete phases shifts, and the resolution achieves $K=10$ different patterns. Fig. \ref{convPerf:Dectionary1} and Fig. \ref{convPerf:Dectionary2} illustrate the ten different patterns used at the final stage in elevation and azimuth , respectively. It should be mentioned that even though channel $\mathbf{G}$ will shift the reflected beam corresponding to each codeword as discussed in Subsection \ref{section:sec2,2}, still the resulted shifted beams will scan the whole space. This is shown by Fig. \ref{convPerf:Dectionary3} and Fig. \ref{convPerf:Dectionary4}, where we can see that the channel $\mathbf{G}$ just caused a rotation in total beam patterns of the codebook. However, the \ac{UE} will use the codebook normally and based on the received optimal codeword, it can find the \ac{AoA} by just comparing this codeword to a predefined table, or from the polar diagram illustrated in Fig. \ref{convPerf:Dectionary1} and Fig. \ref{convPerf:Dectionary2}.    

In the second step, channel $\mathbf{H}$ is estimated using the iterative resolution algorithm and the performance of this algorithm is evaluated using \ac{NMSE} given by
\begin{equation}
    \operatorname{NMSE} = \operatorname{E} \Bigg{[}\frac{||\mathbf{\tilde{H}}-\mathbf{\hat{H}}||^2_F}{||\mathbf{\hat{H}}||^2_F}\Bigg{]}.
\end{equation}
\begin{figure}
    \centering
    \includegraphics[scale=0.42]{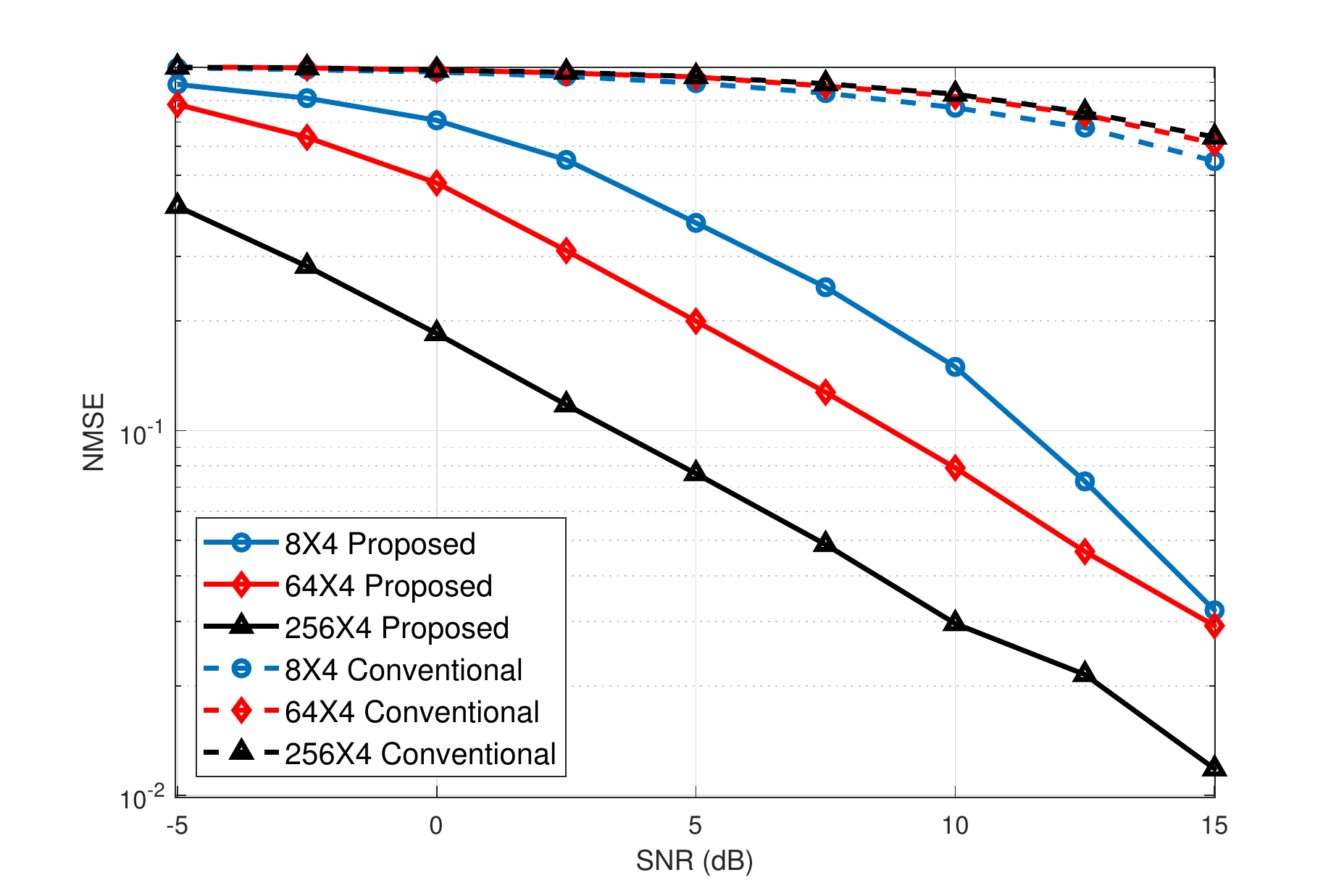}
    \caption{\ac{NMSE} performance comparison between the proposed framework and conventional approaches of channel $\mathbf{H}$ at different \ac{RIS} array sizes. }
    \label{fig:Channel-estimation}
\end{figure}
We consider narrowband mmWave channel with a \ac{MIMO} system, number of antennas at the base station and at \ac{UE} is $N_{BS}=16\times16=256$ and $N_{UE}=2\times 2=4$, respectively. The path gains are assumed to have Gaussian distribution. \ac{RIS} is assumed to have different geometries $M_{RIS}=4\times4=16$, $M_{RIS}=8\times8=64$ and $M_{RIS}=16\times16=256$, for each case number of pilots is $U_p = 8$, $U_p = 32$ and $U_p = 128$, respectively. We also assume one dominant \ac{LoS} path $L_h = 1$ between \ac{RIS}-\ac{UE}. Conventional methods i.e., cascaded channel estimation, were adopted in our system to be compared with the proposed algorithm, and most importantly no prior knowledge of the \ac{UE}'s location is assumed. 

Fig. \ref{fig:Channel-estimation} compares the \ac{NMSE} performance against \ac{SNR}. The proposed scheme achieves very high performance compared the the conventional ones, where the \ac{NMSE} keeps decreasing with the \ac{SNR} increasing, reaching almost $0.01$ normalized error at \ac{SNR}$~=15$ dB and $M_{RIS}=256$. The reason is that in the proposed scheme, the power is focused on the target \ac{UE} before staring channel estimation protocol, this result is also reflected in the same figure at low \ac{SNR} values, where \ac{NMSE} is still relatively small even though the transmit power is minimal. Also, increasing the number \ac{RIS} elements gives the better channel estimation, where the lowest \ac{NMSE} corresponded to $M_{RIS}=256$, then $M_{RIS}=64$, and lastly to $M_{RIS}=16$. However, the conventional algorithms, regardless of the number of \ac{RIS} elements, have very bad \ac{NMSE} performance through all \ac{SNR} values. The reason is that the energy is not beamformed toward the \ac{UE} most of the time, and the received power from the beam sides is always weak, and thus the channel cannot be estimated reliably. The slight change at high \ac{SNR} indicates that the \ac{UE} in this case is receiving slightly higher power, but still it is not enough because it does not directly come from the main beam.  

\subsection{The Channel Tracking Performance}
\begin{figure}
    \centering
    \includegraphics[scale=0.47]{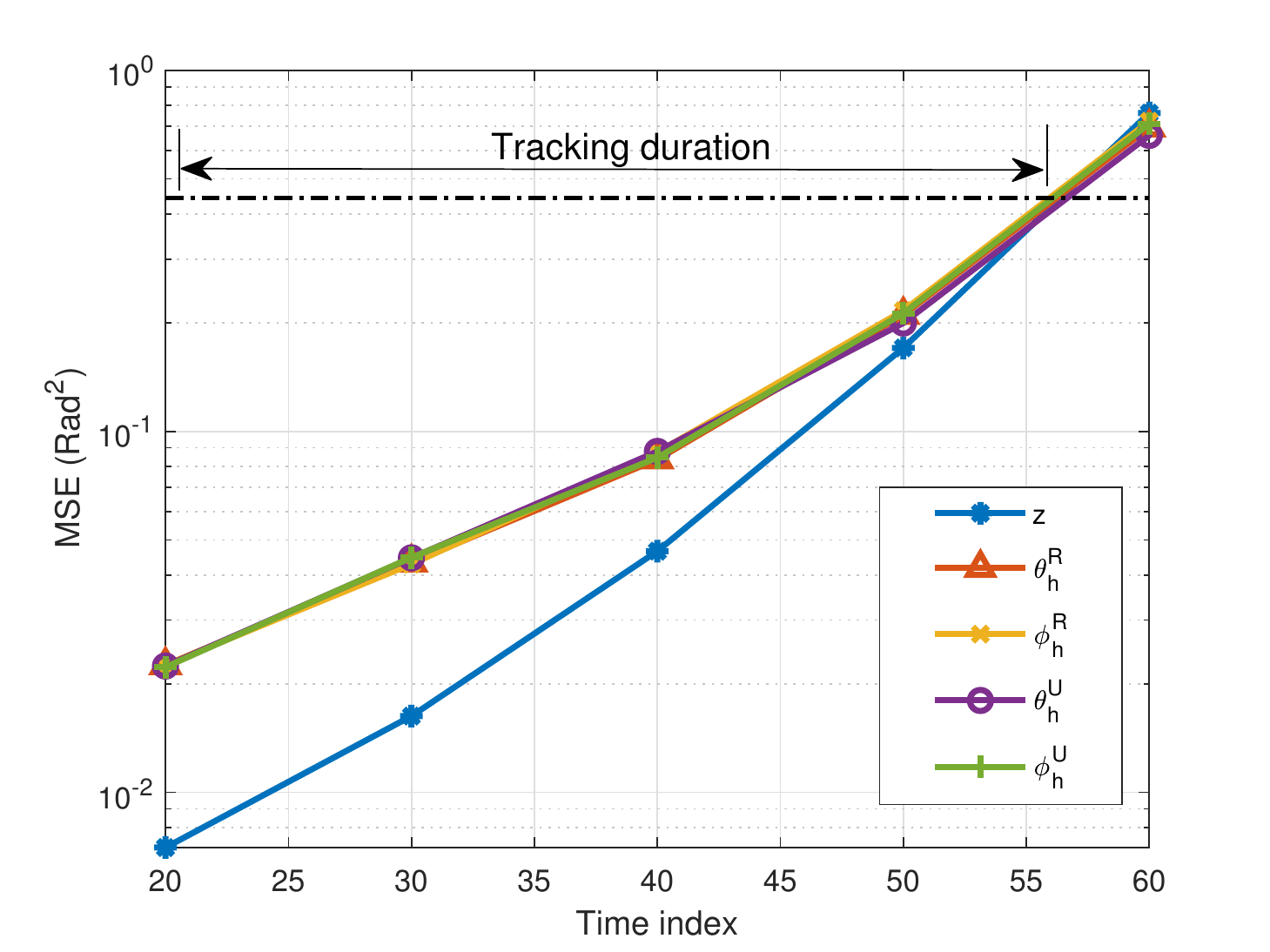}
    \caption{MSE of the tracked parameters: complex path coefficient, elevation and azimuth  \ac{AoD}s from the \ac{RIS}, and elevation and azimuth  \ac{AoA}s at \ac{UE} using \ac{EKF} tracking algorithm.}
    \label{fig:MSEtrack}
\end{figure}

In this subsection, we assume that the channel between the \ac{BS} and \ac{RIS} is fixed and its parameters are constant during the tracking period. Also, it is assumed that the channel remains stationary during this observation interval, and the channel sparsity in the mmWave makes the paths to be likely separated from each other under the assumption that the \ac{RIS} is located near to \ac{UE} location. Hence, only one single path falls into the main beam direction $L_h=1$ \cite{Tr2}, giving that the state space vector at each time index is
\begin{equation} \label{equ:statespacevector}
  \boldsymbol{x}_{state} =  [z_{\Re} ~ z_{\Im} ~ \theta^R_{h} ~ \varphi^R_{h} ~ \theta^U_{h} ~ \varphi^U_{h}]^T,
\end{equation}
where $z = z_{\Re} + jz_{\Im}$. By using the real and imaginary part of $z$, the state vector $\boldsymbol{x}_{state}$ is a real vector which helps to avoid implementation issues when real and complex numbers are combined.  
Fig. \ref{fig:MSEtrack} shows the \ac{MSE} of the tracked channel parameters between \ac{RIS} and \ac{UE} using \ac{EKF} algorithm with the same filter setups as in \cite{Tr2} at \ac{SNR}$=20$ dB. It is clearly shown that the algorithm has the ability to reduce the estimation overhead for longer time since it can keep the error below a certain threshold of half power beamwidth where it is given as $\triangle \theta_{3dB}  \approx  \frac{\lambda}{\sqrt{M_{\operatorname{RIS}}}d} 0.886$ \cite{BW}. 
 
\begin{figure}
    \centering
    \includegraphics[scale=0.47]{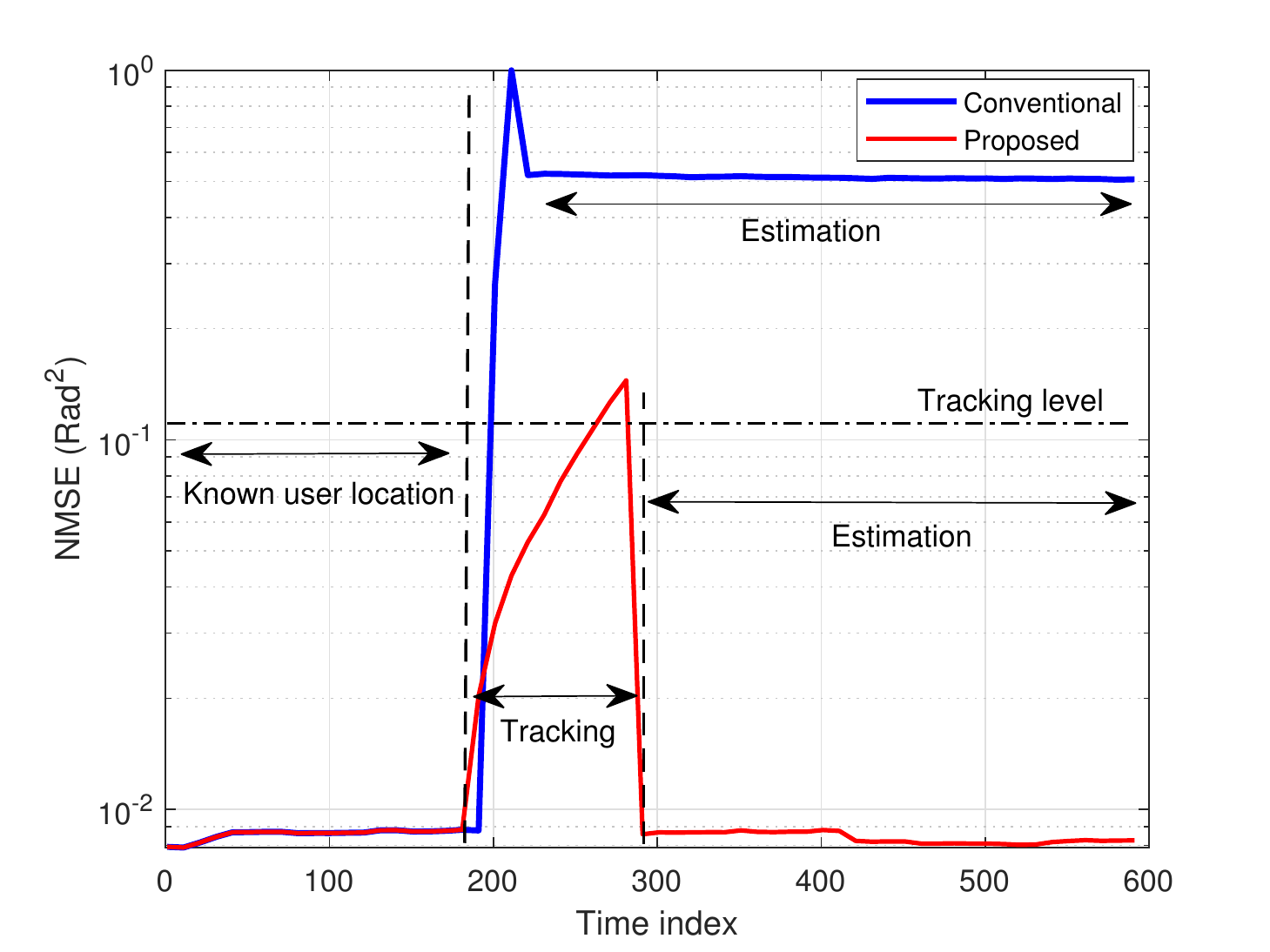}
    \caption{The overall system performance for the proposed three-stage \ac{RIS} framework compared to conventional cascaded channel estimation methods.}
    \label{fig:wholeperformance}
\end{figure}

The overall performance of the proposed three-stage \ac{RIS} framework is illustrated in Fig. \ref{fig:wholeperformance}, where it is assumed that both proposed and conventional RIS-assisted communication channel estimation techniques know the location of the \ac{UE} at the initial state, and that they are beamforming toward this direction at \ac{SNR}$=20$ dB. This assumption is favorable to the conventional scheme. The simulation consists of three states, in which the \ac{UE} is stationary at first, then it starts moving and finally becomes stationary again. Fig. \ref{fig:wholeperformance} shows that at the initial state both methods perform well, achieving low normalized error. However, the performance totally changes as the \ac{UE} moves. In the proposed channel estimation scheme, as the \ac{UE} starts moving, the channel is tracked until a certain threshold and the beamforming is shifted based on the tracked parameters. After that, channel estimation is needed where the result converges again to a minimum \ac{NMSE}. In case of the conventional method, the error increases very fast which needs to be compensated by estimating the channel where the result settles to $\operatorname{NMSE} = 0.5$, resulting in a huge performance gap between the two schemes.

\section{Conclusion} \label{section:sec7}
In this paper, we propose a three-stage framework for an \ac{RIS}-aided mmWave \ac{MIMO} communication system. In the first and the second stages, the channel estimation problem is extensively studied and new approach is proposed to estimate \ac{BS}-\ac{RIS} channel (channel $\mathbf{G}$) and \ac{RIS}-\ac{UE} channel (channel $\mathbf{H}$) separately, by exploiting the shift in the direction of the beams reflected from \ac{RIS} due to $\mathbf{G}$ effect. The estimation of channel $\mathbf{G}$ is used to develop a low-complex, real-time applicable phase design.
Then, the channel $\mathbf{H}$ is estimated using the iterative resolution method and the prior knowledge of \ac{AoD}/\ac{AoA} that were estimated in the previous step. 
In the third stage, channel tracking algorithms are applied to track the channel between \ac{RIS} and \ac{UE}, since estimating channels $\mathbf{G}$ and $\mathbf{H}$ are done separately which allows the user to have some level of mobility. 
The performance analysis showed that the proposed framework can provide an accurate channel estimation.
The proposed framework for RIS-aided communication was developed under very practical assumptions which makes it very useful to be implemented with the available \ac{RIS} prototypes such \textbf{MIT’s RFocus prototype} \cite{rfocus}, that beamforms and focuses the impinging radio waves towards specified direction and location, respectively.

\section*{Acknowledgment}
 This work was supported by the Scientific and Technological Research Council of Turkey under Grant No. 5200030.

\begin{appendices}
\section{Proof of equation \eqref{equ:Theta_design}}
To cancel the effect of channel $\hat{\mathbf{G}}$, the outcome of the effective channel should be given as
\begin{equation}
    \mathbf{\hat{H}\Theta\hat{G}}=\mathbf{\hat{H} \Theta}(\theta_{des},\varphi_{des}) \mathbf{G}_{opt},
\end{equation}
which is equivalent to $\mathbf{\Theta\hat{G}}=\mathbf{ \Theta}(\theta_{des},\varphi_{des}) \mathbf{G}_{opt}$. Then we find
\begin{equation}
\begin{aligned}
    \mathbf{\Theta\hat{G}}&=\mathbf{ \Theta}(\theta_{des},\varphi_{des}) \mathbf{G}_{opt}\\
    \mathbf{\Theta\hat{G}}\mathbf{\hat{G}^H}&=\mathbf{ \Theta}(\theta_{des},\varphi_{des}) \mathbf{G}_{opt}\mathbf{\hat{G}^H}\\
    \mathbf{\Theta}(\hat{\mathbf{G}}\mathbf{\hat{G}^H})(\hat{\mathbf{G}}\mathbf{\hat{G}^H})^{-1}&=\mathbf{ \Theta}(\theta_{des},\varphi_{des}) \mathbf{G}_{opt}\mathbf{\hat{G}^H}(\hat{\mathbf{G}}\mathbf{\hat{G}^H})^{-1}.
\end{aligned}
\end{equation}
From this point, we get \eqref{equ:Theta_design}.
\end{appendices}

\balance

\ifCLASSOPTIONcaptionsoff
  \newpage
\fi

\vfill

\end{document}